\documentclass[article]{jss}

\usepackage{amsmath,amssymb,array,placeins,orcidlink,thumbpdf,float}

\newcommand{\class}[1]{`\code{#1}'}
\newcommand{\fct}[1]{\code{#1()}}
\author{
  Junhao Gao~\orcidlink{0009-0006-5375-190X}\\University of California, San Diego
  \And Ery Arias-Castro~\orcidlink{0000-0002-3038-5736}\\University of California, San Diego
}
\Plainauthor{Junhao Gao, Ery Arias-Castro}

\title{Kernelized Stein Discrepancy for Goodness-of-Fit Tests and Stein Sampling in \proglang{R}}
\Plaintitle{Kernelized Stein Discrepancy for Goodness-of-Fit Tests and Stein Sampling in R}
\Shorttitle{Kernel Goodness-of-Fit Tests and Stein Sampling in \proglang{R}}

\Abstract{
  Stein's method constructs computable discrepancies between a
  target distribution and a candidate distribution without requiring the
  target distribution's normalizing constant.  These discrepancies
  support goodness-of-fit tests for model assessment as well as sampling tools
  for empirical approximation.  The \proglang{R} package
  \pkg{steinsampling} provides the first unified \proglang{R} workflow for
  applying score-based Stein methods to kernel goodness-of-fit testing of
  independent or serially dependent observations, point transport,
  greedy point construction, and sample compression.
  High-level functions carry out each task in a single call, while the kernel,
  calibration, optimization, and transition components are provided separately
  so that users can replace any one of them.  A single score and kernel setup
  can therefore be reused across sampling and testing, making these methods easier
  to reproduce, compare, and extend.
}

\Plainkeywords{kernelized Stein discrepancy, goodness-of-fit testing, Stein's
  method, Stein sampling, sample compression, R}

\Address{
  Junhao Gao\\
  Department of Mathematics\\
  University of California, San Diego\\
  9500 Gilman Drive \# 0112\\
  La Jolla, CA 92093-0112, United States of America\\
  E-mail: \email{jug049@ucsd.edu}\\[1ex]
  Ery Arias-Castro\\
  Department of Mathematics\\
  \emph{and}\\
  Halıcıoğlu Data Science Institute\\
  University of California, San Diego\\
  9500 Gilman Drive \# 0112\\
  La Jolla, CA 92093-0112, United States of America\\
  E-mail: \email{eariascastro@ucsd.edu}
}

\begin{document}

\shortcites{Anastasiou+EtAl:2023, Liu+Lee+Jordan:2016,
  Chwialkowski+Strathmann+Gretton:2016,
  Chwialkowski+EtAl:software,
  Oates+Girolami+Chopin:2017,
  Jitkrittum+Xu+Szabo+Fukumizu+Gretton:2017,
  Jitkrittum+EtAl:kgof,
  Chen+Mackey+Gorham+Briol+Oates:2018,
  Chen+Barp+Briol+Gorham+Girolami+Mackey+Oates:2019,
  Chen+EtAl:SteinPointsSoftware, Chen+EtAl:SPMCMCSoftware,
  Riabiz+Chen+Cockayne+Swietach+Niederer+Mackey+Oates:2022,
  Riabiz+EtAl:software,
  Korba+EtAl:2021, Li+Dwivedi+Mackey:2024, Benard+Staber+DaVeiga:2023,
  Huggins+Mackey:2018,
  Pyro:software, NumPyro:software,
  Harris+EtAl:2020, Pedregosa+EtAl:2011,
  mvtnorm, Roshan+EtAl:2015}
\shortcites{RcppParallel,torch,Schrab+Kim+Guedj+Gretton:2022}



\section{Introduction} \label{sec:intro}

Many target distributions are specified through an unnormalized density, so
probabilities cannot be evaluated directly and exact simulation is often
unavailable.  This raises two statistical problems: testing whether the
distribution of observed data agrees with the target distribution, and
constructing a point set whose empirical distribution approximates it, possibly
by compressing an existing sample.  Stein's method links these problems by
using the score of the target density to define discrepancies that detect
mismatch in the distribution and provide objectives for point transport, point
construction,
and sample selection.

Kernelized Stein discrepancy (KSD)
\citep{Liu+Lee+Jordan:2016,Chwialkowski+Strathmann+Gretton:2016} and finite-set
Stein discrepancy (FSSD) \citep{Jitkrittum+Xu+Szabo+Fukumizu+Gretton:2017}
provide complementary goodness-of-fit tests for evaluating agreement between a
sample and the target distribution.  The same score and kernel setup also
supports methods that transport, construct, or select representative points.
Stein variational gradient descent (SVGD) transports an initial point set
\citep{Liu+Wang:2016}.  Stein Points constructs a deterministic point set to
approximate the target distribution
\citep{Chen+Mackey+Gorham+Briol+Oates:2018}; Stein Point Markov chain Monte
Carlo (SP-MCMC) uses Markov chain transitions for the same goal
\citep{Chen+Barp+Briol+Gorham+Girolami+Mackey+Oates:2019}; and Stein Thinning
compresses an existing sample into a smaller representative subset
\citep{Riabiz+Chen+Cockayne+Swietach+Niederer+Mackey+Oates:2022}.  All share the
score of the target density and a kernel as their main target-specific inputs.

Existing software for Stein methods is fragmented across languages and usually
focuses on one task.  For goodness-of-fit
testing, the \proglang{R} package \pkg{KSD},
now archived on the Comprehensive \proglang{R} Archive Network (CRAN),
implements the U-statistic KSD for independent observations
\citep{Kang+Liu:2021}.  The \proglang{R} package \pkg{MDgof} includes a KSD
statistic within a general simulation-based multivariate goodness-of-fit suite
\citep{MDgof}.  The \proglang{Python} package \pkg{kgof} implements FSSD.  The
\proglang{Python} code \code{kernel\_goodness\_of\_fit} implements a
V-statistic KSD.  Both reproduce methods from the original articles
\citep{Jitkrittum+EtAl:kgof,Chwialkowski+EtAl:software}.  The latter is
described by its authors as experimental and potentially incomplete.  The
\proglang{Julia}
package \pkg{SteinDiscrepancy.jl} implements a V-statistic KSD
\citep{SteinDiscrepancy:software}.  The \proglang{Julia} package
\pkg{KernelGoodnessOfFit.jl} implements a U-statistic KSD and FSSD, with an
open TODO for its KSD interface
\citep{KernelGoodnessOfFit:software}.

For point transport, the \proglang{Python} libraries \pkg{Pyro} and
\pkg{NumPyro} provide SVGD but no other method
\citep{Pyro:software,NumPyro:software}.  The \proglang{Python} package
\pkg{ksddescent} also provides an SVGD option \citep{Korba+EtAl:2021}.
Reference implementations of SVGD are available in \proglang{Python} and
\proglang{MATLAB} \citep{Liu+Wang:software}.  Stein Points and SP-MCMC
are provided in \proglang{MATLAB}
\citep{Chen+EtAl:SteinPointsSoftware,Chen+EtAl:SPMCMCSoftware}.  Stein
Thinning is available in \proglang{R}, \proglang{Python}, and \proglang{MATLAB}
\citep{Riabiz+Chen+Cockayne+Swietach+Niederer+Mackey+Oates:2022,Riabiz+EtAl:software}.  For thinning, the
\proglang{Python}/\proglang{JAX} packages \pkg{goodpoints} and \pkg{kernax}
provide Stein Thinning among other options
\citep{Li+Dwivedi+Mackey:2024,Benard+Staber+DaVeiga:2023}.  These
implementations do not provide a unified workflow for Stein sampling and
goodness-of-fit testing.

\pkg{steinsampling} is, to our knowledge, the first \proglang{R} package to
unify KSD and FSSD goodness-of-fit testing with SVGD, Stein Points, SP-MCMC,
and Stein Thinning in a single workflow spanning setup, testing, sampling, and
compression.  These stages are deliberately decoupled.
Each can be called independently, replaced by an alternative implementation,
or combined into a custom workflow.  The package provides both high-level
algorithms and lower-level helper functions for this purpose.  Kernels,
optimizers, bootstrap or calibration procedures, transition mechanisms, and
other computational components can be selected or replaced without changing the
surrounding workflow.  The emphasis is therefore on breadth and
interoperability rather than throughput.  The package is pure \proglang{R}
with no compiled, parallel, or GPU backend.  For a single method at large
sample sizes, the \proglang{JAX} implementations in \pkg{goodpoints} and
\pkg{kernax} are the better choice.

Section~\ref{sec:stein} introduces Stein's method,
Sections~\ref{sec:gof} and~\ref{sec:sampling} present goodness-of-fit testing
and Stein sampling and sample compression, respectively,
Section~\ref{sec:application} gives case studies, and
Section~\ref{sec:summary} concludes.



\section{Stein's method in brief} \label{sec:stein}

Stein's method compares a candidate distribution $Q$ with a target
distribution $P$, with density $p$ on $\mathbb{R}^d$, through expectation
identities that are specific to $P$ \citep{Stein:1972}. The
corresponding expectations equal zero under $P$, as expressed by the Stein identity
\begin{equation} \label{eq:stein-identity}
\E_{X\sim P}\{(\mathcal T_p f)(X)\}=0,
\qquad f\in\mathcal G_p.
\end{equation}
Here $\mathcal G_p$ denotes a class of admissible test functions for which the
identity holds.  For the Langevin--Stein operator below, $f$ is a vector field.
The Stein operator $\mathcal T_p$ maps $f$ to the scalar function $\mathcal T_p f$.
Section~\ref{sec:operator} gives the conditions this operator requires
\citep{Anastasiou+EtAl:2023}.

The identity provides the basis for a discrepancy between $Q$ and $P$. For a
sufficiently rich class of test functions, a departure from $P$ can be
detected through a nonzero Stein expectation. The supremum of the absolute
deviation over a suitably constrained class
$\mathcal F\subseteq\mathcal G_p$ for which the expectations exist,
\begin{equation} \label{eq:stein-discrepancy}
\mathcal D(Q,P)
=\sup_{f\in\mathcal F}
\left|\E_{X\sim Q}\{(\mathcal T_p f)(X)\}\right|,
\end{equation}
defines a Stein discrepancy for comparing $Q$ with $P$
\citep[Section~3]{Gorham+Mackey:2015}. Two choices remain open: the
operator $\mathcal T_p$ and the class $\mathcal F$.
Section~\ref{sec:operator} constructs an operator that can be evaluated
from the unnormalized density, and Section~\ref{sec:kernelized} chooses
$\mathcal F$ so that $\mathcal D(Q,P)$ becomes an explicit formula.

\subsection{The Langevin--Stein operator and score function}
\label{sec:operator}

Assume that $P$ admits a continuously differentiable, strictly positive density $p$ on $\mathbb{R}^d$. For a
differentiable vector field
$f=(f_1,\ldots,f_d)^\top:\mathbb{R}^d\to\mathbb{R}^d$, assume the required regularity, integrability, and tail conditions. The
divergence theorem \citep[Section~2.2.2]{Oates+Girolami+Chopin:2017} gives
$\int_{\mathbb{R}^d}\nabla\!\cdot\{p(x)f(x)\}\,dx=0$. The divergence
expands as
\begin{equation*}
\nabla\!\cdot\{p(x)f(x)\}
=p(x)\left\{\nabla\log p(x)^\top f(x)+\nabla\!\cdot f(x)\right\},
\qquad
\nabla\!\cdot f(x)
=\sum_{r=1}^d\frac{\partial f_r(x)}{\partial x_r}.
\end{equation*}
Thus the expression in braces has expectation zero under $P$ and defines the
Langevin--Stein operator $\mathcal T_p$
\citep{Gorham+Mackey:2015}:
\begin{equation} \label{eq:langevin-stein-operator}
(\mathcal T_p f)(x)
=s_p(x)^\top f(x)+\nabla\!\cdot f(x),
\qquad
s_p(x):=\nabla\log p(x),
\end{equation}
where $s_p$ is the score function of the density $p$. Under the preceding
regularity conditions, $\mathcal T_p$ satisfies the Stein identity in
Equation~\ref{eq:stein-identity}.
The implementation therefore applies directly to continuous densities on
$\mathbb R^d$ under these conditions.  It does not apply, for example, to
discrete distributions, which require a different Stein operator.

The operator is evaluable from an unnormalized density because it depends on
$p$ only through its score, and the score does not depend on the normalizing
constant. This settles the first of the two choices left open by
Equation~\ref{eq:stein-discrepancy}, the operator $\mathcal T_p$,
and Section~\ref{sec:kernelized} settles the second, the class $\mathcal F$.

\subsection{Kernels and reproducing kernel Hilbert space}
\label{sec:kernelized}

Let $k:\mathbb R^d\times\mathbb R^d\to\mathbb R$ be a twice continuously
differentiable positive-definite base kernel whose reproducing kernel Hilbert
space (RKHS) is $\mathcal H_k$, and
let $\mathcal H_k^d$ denote the space of vector fields whose components belong
to $\mathcal H_k$. We take $\mathcal F$ to be the unit ball of
$\mathcal H_k^d$, following
\citet[Section~2.1]{Chwialkowski+Strathmann+Gretton:2016},
\citet[Equation~(13)]{Liu+Lee+Jordan:2016}, and
\citet[Section~3.1]{Gorham+Mackey:2017}. We assume that
$\mathcal H_k^d\subseteq\mathcal G_p$, so every vector field considered here
satisfies the Stein identity in Equation~\ref{eq:stein-identity}.

This choice makes Equation~\ref{eq:stein-discrepancy} tractable because the
reproducing property expresses operator evaluations as RKHS inner products,
and the supremum of an inner product over the unit ball is the corresponding
RKHS norm. In particular,
$g(x)=\langle g,k(x,\cdot)\rangle_{\mathcal H_k}$, and, because $k$ is smooth,
derivatives of $g$ are represented by the corresponding derivatives of
$k(x,\cdot)$. Applying these facts to the score and divergence
terms in Equation~\ref{eq:langevin-stein-operator} gives
\begin{equation} \label{eq:stein-feature}
(\mathcal T_p f)(x)
=\left\langle f,\xi_p(x,\cdot)\right\rangle_{\mathcal H_k^d},
\qquad
\xi_p(x,\cdot)
=s_p(x)k(x,\cdot)+\nabla_x k(x,\cdot).
\end{equation}
To compute the resulting RKHS norm, we use the closed-form inner product
between $\xi_p(x,\cdot)$ and $\xi_p(y,\cdot)$, which defines the pairwise
Stein kernel $k_{0,p}$
\citep{Liu+Lee+Jordan:2016,Chwialkowski+Strathmann+Gretton:2016,Anastasiou+EtAl:2023}:
\begin{align} \label{eq:stein-kernel}
k_{0,p}(x,y)
:={}&\left\langle
\xi_p(x,\cdot),\xi_p(y,\cdot)
\right\rangle_{\mathcal H_k^d} \notag\\
={}&s_p(x)^\top s_p(y)k(x,y)
+s_p(x)^\top\nabla_y k(x,y) \notag\\
&+s_p(y)^\top\nabla_x k(x,y)
+\operatorname{tr}\!\left\{\nabla_x\nabla_y^\top k(x,y)\right\}.
\end{align}
When $\E_{X\sim Q}\{k_{0,p}(X,X)\}$ is finite and $X,X'\sim Q$ are independent
\citep[Theorem~2.1]{Chwialkowski+Strathmann+Gretton:2016}, the squared KSD has
the following representation.
\begin{equation} \label{eq:ksd}
\mathrm{KSD}_{k,p}^2(Q)
=\E_{X,X'\sim Q}\{k_{0,p}(X,X')\}.
\end{equation}
Under suitable regularity, integrability, and kernel conditions,
$\mathrm{KSD}_{k,p}(Q)=0$ implies $Q=P$
\citep[Theorem~2.2]{Chwialkowski+Strathmann+Gretton:2016}.

The two built-in base kernels are the radial basis function (RBF) kernel,
also known as the Gaussian kernel, and the inverse multiquadric (IMQ) kernel:
\begin{align}
k_{\mathrm{RBF}}(x,y)
&=\exp\!\left\{-\frac{\lVert x-y\rVert^2}{2h^2}\right\},
&&h>0, \label{eq:rbf-kernel}\\
k_{\mathrm{IMQ}}(x,y)
&=\{c^2+\lVert x-y\rVert^2\}^{\beta},
&&c>0,\quad\beta<0. \label{eq:imq-kernel}
\end{align}
Here $h$ and $c$ set the RBF and IMQ distance scales, respectively, and
$\beta$ controls the IMQ decay.  The RBF kernel decays exponentially and the
IMQ kernel polynomially. For a sequence of distributions $Q_n$, whether
$\mathrm{KSD}_{k,p}(Q_n)\to0$ implies $Q_n\Rightarrow P$ depends on the kernel.
Under the conditions of \citet[Theorem~8]{Gorham+Mackey:2017}, this implication
holds for the IMQ kernel with $\beta\in(-1,0)$, but for the RBF kernel it need not
hold \citep[Theorem~6]{Gorham+Mackey:2017}.  Section~\ref{sec:gof-kernels}
describes the kernel setup.



\section{Goodness-of-fit testing} \label{sec:gof}

Let $X_1,\ldots,X_n$ have common marginal distribution $Q$, and let $P$ be a
fixed target distribution whose score $s_p$ can be evaluated.  The
goodness-of-fit problem considered here is
\begin{equation} \label{eq:gof-hypotheses}
H_0:Q=P
\qquad\text{against}\qquad
H_1:Q\ne P.
\end{equation}
All tests below compare a fixed $P$ with observations from $Q$ using
the Stein feature in Equation~\ref{eq:stein-feature}, without sampling from
$P$ or normalizing its density.
Table~\ref{tab:gof-tests} summarizes the
components of each goodness-of-fit test.  At significance level $\alpha$, each
test rejects $H_0$ when its observed statistic exceeds the $(1-\alpha)$-quantile
of its null approximation.  For FSSD-rand, this is a same-sample
plug-in rule rather than a guaranteed level-$\alpha$ calibration.
Section~\ref{sec:gof-interface}
introduces the common setup, Section~\ref{sec:gof-kernels} the kernel
setup, and Sections~\ref{sec:ksd-tests} and~\ref{sec:fssd} the KSD and
FSSD tests.

\begin{table}[!ht]
\centering
\small
\begin{tabular}{p{0.13\linewidth}p{0.20\linewidth}p{0.21\linewidth}p{0.35\linewidth}}
\hline
Method & Observations & Reported statistic & Null approximation \\
\hline
KSD-U & i.i.d.\ & $nU_n$ (no diagonal) &
centered multinomial bootstrap \\
KSD-V & i.i.d.\ & $nV_n$ (with diagonal) &
Rademacher wild bootstrap \\
KSD-V & serially dependent & $nV_n$ (with diagonal) &
Markov wild bootstrap \\
FSSD-rand & i.i.d.\ & $n\widehat{\mathrm{FSSD}}^2$ &
same-sample plug-in law (heuristic) \\
FSSD-opt & i.i.d.\ & $n_{\mathrm{test}}\widehat{\mathrm{FSSD}}^2$ &
the same null law on held-out rows \\
\hline
\end{tabular}
\caption{\label{tab:gof-tests}Goodness-of-fit statistics and null
approximations provided by \pkg{steinsampling}.}
\end{table}

\subsection{Common setup} \label{sec:gof-interface}

The examples below use \pkg{steinsampling}~0.1.3 under
\proglang{R}~4.6.0
\citep{R}.
The package is available from CRAN at
\url{https://CRAN.R-project.org/package=steinsampling}.
All code shown in Sections~\ref{sec:gof}--\ref{sec:application} is collected in the replication script,
which also defines the Section~\ref{sec:application} functions that are not part of the package.
Every goodness-of-fit routine in this section takes three core inputs:
observations in an $n\times d$ numeric matrix \code{X}, a vectorized
\code{score_function} mapping a matrix of $d$-dimensional rows to a matrix of
the same shape with rows $s_p(x_i)^\top$, and a base kernel.  Because the stated null
calibrations treat the density $p$ and its score as fixed, any estimated
parameters of $p$ must be based on data independent of \code{X}.

The following target distribution, its score, and the observations are used
throughout the section:
\begin{CodeChunk}
\begin{CodeInput}
R> library("steinsampling")
R> target <- gmm(nComp = 1, mu = c(0, 0), sigma = diag(2), d = 2)
R> score_p <- get_score_evaluator(target)
R> set.seed(2026)
R> X <- rgmm(target, n = 400)
\end{CodeInput}
\end{CodeChunk}
\code{target} represents $\mathcal{N}_2(0,I_2)$, and \fct{rgmm} draws the
$400$ independent rows of \code{X} from it.  The package's multivariate
Gaussian density and simulation routines use \pkg{mvtnorm}
\citep{mvtnorm,Genz+Bretz:2009}.
\fct{get\_score\_evaluator} returns the score function $s_p$ as
\code{score_p}.  For this Gaussian density,
\code{score_p(X)} is the $400\times2$ matrix \code{-X}.
Users can also supply functions from other packages or define their own for
more complex or customized setups.
The same \code{X} and \code{score_p} are passed to every test below,
except that Section~\ref{sec:ksd-v} changes the observations to illustrate dependent
data. 

\subsection{Kernel setup} \label{sec:gof-kernels}

The score function and the base kernel are supplied separately.  A
\class{SteinKernel} object stores the base kernel $k$ and its parameters.
The function \fct{stein\_kernel} creates the built-in RBF and IMQ kernels.
The test functions accept either a \class{SteinKernel} object or the name
\code{"gaussian_rbf"} or \code{"imq"}.

The optional positive-definite matrix \code{precon} changes the squared
distance from $\lVert x-y\rVert^2$ to
$(x-y)^\top\mathtt{precon}(x-y)$.  If \code{precon} is not supplied, the
ordinary squared Euclidean distance is used.  Given the score values and the
kernel, \fct{stein\_kernel\_matrix} computes
$K_{ij}=k_{0,p}(X_i,X_j)$ from Equation~\ref{eq:stein-kernel}.

The running example uses a fixed RBF kernel:
\begin{CodeChunk}
\begin{CodeInput}
R> gof_kernel <- stein_kernel("gaussian_rbf", h = 1)
\end{CodeInput}
\end{CodeChunk}
For built-in kernels, \code{scaling} denotes the squared scale: $h^2$ for RBF
and $c^2$ for IMQ.  If a kernel object stores a fixed scale, the test uses that
value.  KSD and FSSD-rand keep the value fixed.  FSSD-opt uses it as a starting
value and may update it.

If the kernel object does not store a fixed scale, the test uses the supplied
\code{scaling}.  When \code{scaling = NULL}, KSD and FSSD-rand compute the
median scale from all input rows.  FSSD-opt instead computes it from the
training rows.  Thus, \code{gof_kernel} fixes \code{scaling = 1} for KSD and
FSSD-rand and initializes FSSD-opt at $1$.

Users can define a custom symmetric base kernel with:
\begin{Code}
custom_stein_kernel(eval, grad_x, trace_mixed,
                    fssd_grad = NULL, scale2 = NULL, precon = NULL,
                    custom_grad_mode = c("analytic", "numeric"))
\end{Code}
The arguments \code{eval}, \code{grad_x}, and \code{trace_mixed} accept
user-supplied functions for the kernel value, its gradient, and its mixed
derivative trace.  Each supplied function receives \code{k}, \code{X}, \code{Y}, and
\code{M}, where \code{M} is the preconditioning matrix supplied through
\code{precon}, or \code{NULL} if none is supplied.  The package passes
\code{M} to these functions but does not automatically apply it, so any use of
\code{M} must be implemented in the supplied functions.

For point matrices $X\in\mathbb R^{n_X\times d}$ and
$Y\in\mathbb R^{n_Y\times d}$, \fct{eval\_kernel} calls the function
supplied through \code{eval} and returns the $n_X\times n_Y$ kernel matrix
$k(X,Y)$.
\fct{grad\_x\_kernel} calls the function supplied through \code{grad_x}
and returns the $n_X\times n_Y\times d$ array $\nabla_x k(X,Y)$.
\fct{trace\_mixed\_kernel} calls the function supplied through
\code{trace_mixed} and returns the $n_X\times n_Y$ matrix
$\operatorname{tr}\{\nabla_x\nabla_y^\top k(X,Y)\}$.
\fct{cross\_kernel} computes the sum of the second and third terms in
Equation~\ref{eq:stein-kernel}.  \fct{stein\_kernel\_matrix} combines these quantities to form the
complete Stein-kernel matrix.

KSD uses all three functions.  FSSD uses only \code{eval} and
\code{grad_x}.  For FSSD-opt, a custom kernel must supply a gradient function
through \code{fssd_grad}, evaluated by \fct{fssd\_grad\_kernel},
or use \code{custom_grad_mode = "numeric"}.
The optional argument \code{scale2} sets the custom kernel's initial squared
scale.  \fct{scale2\_kernel} reads or changes this value.  To let \mbox{FSSD-opt}
optimize the scale, supply a function through \code{fssd_grad} that computes
the required gradients.  With \code{custom_grad_mode = "numeric"}, only the
test locations can be optimized.
Section~\ref{sec:fssd-opt} describes the optimization.
For custom kernels, check the Stein conditions in Section~\ref{sec:stein}
and the assumptions of the chosen method in Sections~\ref{sec:gof}
and~\ref{sec:sampling}.

\subsection{Kernelized Stein discrepancy tests} \label{sec:ksd-tests}

The KSD tests assess whether observations from $Q$ agree with the fixed $P$.
Both tests use the pairwise Stein-kernel matrix
$K_{ij}=k_{0,p}(X_i,X_j)$, constructed using the kernel setup in
Section~\ref{sec:gof-kernels}.  KSD-U leaves out the diagonal entries of $K$.
For independent observations, it is unbiased for the squared population KSD
but may be negative.  KSD-V keeps the diagonal entries.  With a
positive-definite base kernel, it is nonnegative but biased upward.  With the
number of bootstrap draws fixed, both tests cost $O(n^2d)$.  The following
subsections define each statistic, explain its bootstrap calibration, and show
the one-call test and its component functions.

\subsubsection{U-statistic test for independent observations}
\label{sec:ksd-u}

For independent observations, KSD-U removes the diagonal of $K$ and computes
\begin{equation} \label{eq:ksd-u}
U_n
=\frac{1}{n(n-1)}
  \sum_{1\le i\ne j\le n}K_{ij},
\qquad
T_n=nU_n.
\end{equation}
$U_n$ is an unbiased estimator of the squared population KSD in
Equation~\ref{eq:ksd}.  The package follows Equation~\ref{eq:ksd-u}
directly.  \fct{ksd\_uq\_matrix} evaluates
$k_{0,p}(X_i,X_j)$ for every pair and returns $K$.
\fct{ksd\_u\_statistic} then sets $K_{ii}=0$, sums the remaining entries, and
divides by $n-1$ to return $T_n=nU_n$.  Thus \fct{ksd\_u\_statistic} and
\fct{ksd\_u\_test} both report the same test-scale value as \code{ksd_u}.

Under $H_0$ and the conditions of
\citet[Theorems~4.1 and~4.3]{Liu+Lee+Jordan:2016}, the package approximates
the limiting law of $nU_n$ using the centered multinomial bootstrap of
\citet[Equation~(16)]{Liu+Lee+Jordan:2016}.  For bootstrap replication $b$, let
$(N_1^{(b)},\ldots,N_n^{(b)})$ be multinomial counts from $n$ draws with equal
cell probabilities and define
\begin{equation} \label{eq:ksd-u-bootstrap}
w_i^{(b)}=\frac{N_i^{(b)}}{n}-\frac{1}{n},
\qquad
U_n^{*(b)}
=\sum_{1\le i\ne j\le n}w_i^{(b)}w_j^{(b)}K_{ij},
\qquad
T_n^{*(b)}=nU_n^{*(b)}.
\end{equation}
\fct{ksd\_u\_bootstrap} implements Equation~\ref{eq:ksd-u-bootstrap} for a
supplied matrix $K$.
\fct{ksd\_u\_test} performs the matrix, statistic, and bootstrap steps
together and returns the right-tail bootstrap $p$~value obtained by comparing
$T_n$ with the bootstrap draws.

The running example uses the inputs prepared in
Sections~\ref{sec:gof-interface} and~\ref{sec:gof-kernels}:
\begin{CodeChunk}
\begin{CodeInput}
R> set.seed(1101)
R> fit_u <- ksd_u_test(X, score_p, kernel = gof_kernel, nboot = 9999)
R> fit_u
\end{CodeInput}
\begin{CodeOutput}
Kernelized Stein Discrepancy (U-statistics) - gaussian_rbf

data:  X
ksd_u = -1.2785, nboot = 9999, scaling = 1, p-value = 0.803
\end{CodeOutput}
\end{CodeChunk}
The test does not reject $P$ ($T_n=-1.2785$, $p=0.803$).  The U-statistic can
be negative in finite samples.  Serially dependent observations require the
V-statistic test below.

\subsubsection{V-statistic test for independent or serially dependent
  observations}
\label{sec:ksd-v}

The V-statistic retains the diagonal of the same Stein-kernel matrix and
computes
\begin{equation} \label{eq:ksd-v}
V_n
=\frac{1}{n^2}\sum_{i=1}^{n}\sum_{j=1}^{n}K_{ij},
\qquad
nV_n
=\frac{1}{n}\sum_{i=1}^{n}\sum_{j=1}^{n}K_{ij}.
\end{equation}
For a valid positive-definite base kernel, $V_n$ is nonnegative but biased for
the squared population KSD in Equation~\ref{eq:ksd}.
\fct{ksd\_vq\_matrix} returns $K$.  \fct{ksd\_v\_statistic} retains every
entry and divides their sum by $n$, and \fct{ksd\_v\_test} reports the result
$nV_n$ as \code{ksd_v}.

Under $H_0$, the package approximates the limiting law of $nV_n$ with the wild bootstrap of
\citet{Chwialkowski+Strathmann+Gretton:2016}.  Given signs
$W_1^{(b)},\ldots,W_n^{(b)}$, define the bootstrap statistic for replication $b$ by
\begin{equation} \label{eq:ksd-v-bootstrap}
nV_n^{*(b)}
=\frac{1}{n}\sum_{i=1}^{n}\sum_{j=1}^{n}
  W_i^{(b)}W_j^{(b)}K_{ij}.
\end{equation}
\fct{ksd\_v\_bootstrap} implements
Equation~\ref{eq:ksd-v-bootstrap} for a supplied matrix $K$ and returns the
scaled draws $nV_n^{*(b)}$.  \fct{ksd\_v\_test} generates the sign matrix,
compares these draws with $nV_n$, and returns the right-tail bootstrap
$p$~value.

For independent observations, set \code{boot_method = "rademacher"}.
For serially dependent observations,
\code{boot_method = "markov"} instead uses the Markov wild bootstrap,
generating the correlated sign sequence
\begin{equation} \label{eq:ksd-v-markov}
W_1^{(b)}=1,
\qquad
W_t^{(b)}
=
\begin{cases}
-W_{t-1}^{(b)}, & \text{with probability }a_n,\\
\phantom{-}W_{t-1}^{(b)}, & \text{with probability }1-a_n,
\end{cases}
\qquad t=2,\ldots,n.
\end{equation}
Here \code{change_prob} is $a_n$, the probability that the sign changes
between adjacent rows.
For this Markov wild bootstrap to consistently approximate the null law of
$nV_n$, the ordered observations must form a stationary process whose serial
dependence, measured by the $\tau$-mixing coefficients
$\tau_{\mathrm{mix}}(t)$, decays fast enough that
$\sum_{t\geq1}t^2\sqrt{\tau_{\mathrm{mix}}(t)}<\infty$; the Stein kernel
$k_{0,p}$ must be Lipschitz with $\E\{k_{0,p}(Z,Z)^2\}<\infty$; and the bootstrap signs must be
generated independently of the observations
\citep{Chwialkowski+Strathmann+Gretton:2016}.  The flip probability must satisfy
$a_n\to0$ and $na_n\to\infty$, producing longer sign runs but an increasing
number of flips, so \code{change_prob} depends on $n$ rather than being a
universal constant.

For independent observations, use \code{boot_method = "rademacher"}:
\begin{CodeChunk}
\begin{CodeInput}
R> set.seed(1102)
R> fit_v <- ksd_v_test(X, score_p, boot_method = "rademacher",
+    kernel = gof_kernel, nboot = 9999)
R> fit_v
\end{CodeInput}
\begin{CodeOutput}
Kernelized Stein Discrepancy (V-statistics) - rademacher bootstrap,
gaussian_rbf kernel

data:  X
ksd_v = 2.6914, nboot = 9999, scaling = 1, change_prob = NA,
p-value = 0.8015
\end{CodeOutput}
\end{CodeChunk}
The reported \code{ksd_v} is $nV_n=2.6914$, and its $p$~value of $0.8015$ does not
reject $P$.  Relative to KSD-U, this statistic retains
$K_{ii}$ and its bootstrap uses sign weights instead of centered multinomial
weights.

The following example uses a correlated Gaussian sequence whose marginal
distribution remains $P=\mathcal N_2(0,I_2)$:
\begin{equation} \label{eq:dependent-ar1}
Z_1\sim\mathcal N_2(0,I_2),\qquad
\epsilon_t\stackrel{\mathrm{iid}}{\sim}\mathcal N_2(0,I_2),\qquad
Z_t=\phi Z_{t-1}+\sqrt{1-\phi^2}\,\epsilon_t.
\end{equation}
With the innovations independent of $Z_1$, setting $\phi=0.8$ gives
$\COV(Z_t,Z_{t-u})=\phi^u I_2$.  The code simulates this
sequence directly:
\begin{CodeChunk}
\begin{CodeInput}
R> set.seed(1201)
R> phi <- 0.8
R> Z <- matrix(0, 400, 2)
R> Z[1, ] <- rnorm(2)
R> for (t in 2:400)
+    Z[t, ] <- phi * Z[t - 1, ] + sqrt(1 - phi^2) * rnorm(2)
R> set.seed(1202)
R> ksd_v_test(Z, score_p, boot_method = "markov", change_prob = 0.02,
+    kernel = gof_kernel, nboot = 9999)
R> set.seed(1203)
R> ksd_v_test(Z, score_p, boot_method = "rademacher",
+    kernel = gof_kernel, nboot = 9999)
\end{CodeInput}
\begin{CodeOutput}
Kernelized Stein Discrepancy (V-statistics) - markov bootstrap,
gaussian_rbf kernel

data:  Z
ksd_v = 12.855, nboot = 9999.00, scaling = 1.00, change_prob = 0.02,
p-value = 0.396

Kernelized Stein Discrepancy (V-statistics) - rademacher bootstrap,
gaussian_rbf kernel

data:  Z
ksd_v = 12.855, nboot = 9999, scaling = 1, change_prob = NA, p-value =
3e-04
\end{CodeOutput}
\end{CodeChunk}
Both calibrations use $nV_n=12.855$.  In this serially dependent setting, the
Markov wild bootstrap does not reject the correct $P$ ($p=0.396$), whereas
the independent-sign bootstrap does ($p=0.0003$).  For serially dependent
samples, \citet{Chwialkowski+Strathmann+Gretton:2016} recommend choosing the
thinning interval and sign-change probability according to the residual
dependence.

\subsection{Finite-set Stein discrepancy} \label{sec:fssd}

FSSD measures discrepancy through the Stein witness at a finite set of
locations $L=\{v_1,\ldots,v_J\}$.  A nonzero witness at any selected
location provides evidence against $Q=P$:
\begin{equation} \label{eq:fssd-population}
g_{p,Q}(v)
=\E_{X\sim Q}\{\xi_p(X,v)\},
\qquad
\mathrm{FSSD}_{p,k,L}^2(Q)
=\frac{1}{dJ}\sum_{j=1}^{J}
  \lVert g_{p,Q}(v_j)\rVert^2.
\end{equation}
Thus $L$ specifies where the discrepancy is evaluated, and $J$ is the number of
test locations.  Increasing $J$ uses more locations but raises the row-feature
dimension to $dJ$ \citep{Jitkrittum+Xu+Szabo+Fukumizu+Gretton:2017}.
Whereas KSD uses a global RKHS norm, Equation~\ref{eq:fssd-population}
averages squared witness values at these $J$ locations.  To estimate this
quantity, define the row feature
\begin{equation} \label{eq:fssd-tau}
\tau(x)
=\frac{1}{\sqrt{dJ}}\operatorname{vec}
  \left[
    \xi_p(x,v_1),\ldots,\xi_p(x,v_J)
  \right]\in\mathbb{R}^{dJ}.
\end{equation}
That is, $\tau(x)$ collects the Stein feature of one observation at all $J$
locations.  The function \fct{compute\_tau} returns these rows as an
$\tilde n\times dJ$ matrix for the $\tilde n$ observations used in the test.
If $L$ and the kernel, including its
scale, are fixed independently of those rows, then
$\E\{\tau(X)^\top\tau(X')\}=\mathrm{FSSD}_{p,k,L}^2(Q)$ for
independent $X,X'\sim Q$, so averaging over ordered pairs $i\ne j$ estimates
Equation~\ref{eq:fssd-population} without bias:
\begin{equation} \label{eq:fssd-u-statistic}
\widehat{\mathrm{FSSD}}^2
=\frac{1}{\tilde n(\tilde n-1)}
  \sum_{1\leq i\ne j\leq \tilde n}\tau(X_i)^\top\tau(X_j)
=\frac{1}{\tilde n(\tilde n-1)}
  \left\{
    \left\lVert\sum_{i=1}^{\tilde n}\tau(X_i)\right\rVert^2
    -\sum_{i=1}^{\tilde n}\lVert\tau(X_i)\rVert^2
  \right\}.
\end{equation}
\fct{fssd\_statistic} evaluates Equation~\ref{eq:fssd-u-statistic} from that
matrix, the second form requiring no $\tilde n\times\tilde n$ array of pairwise
products, and returns the test-scale statistic
$S_{\tilde n}=\tilde n\widehat{\mathrm{FSSD}}^2$ used by
\fct{fssd\_test} and the null simulator.
For fixed $J$ and supplied scores, this U-statistic
therefore costs $O(\tilde ndJ)$, linear in the number of observations rather
than the $O(\tilde n^2d)$ of the KSD tests in Section~\ref{sec:ksd-tests}.

Under $H_0$, suppose that $L$ and the kernel, including its scale, are fixed
independently of the rows being tested, that those rows are i.i.d.\ from $P$,
and that independent $X,X'\sim P$ satisfy
$\E[\{\tau(X)^\top\tau(X')\}^2]<\infty$.  Because $\tau$ takes values in
$\mathbb{R}^{dJ}$, the limit contains only $dJ$ weights:
\begin{equation} \label{eq:fssd-null}
S_{\tilde n}=\tilde n\widehat{\mathrm{FSSD}}^2
\ \xrightarrow{d}\
\sum_{q=1}^{dJ}\lambda_q(Z_q^2-1),
\end{equation}
with independent standard normal $Z_q$ and weights $\lambda_q$ equal to
the eigenvalues of the feature covariance
$\Sigma_P=\COV_{P}\{\tau(X)\}$
\citep{Jitkrittum+Xu+Szabo+Fukumizu+Gretton:2017}.
By contrast, the KSD limit generally contains infinitely many spectral weights,
and the package calibrates its KSD tests by bootstrap.
Because $\Sigma_P$ is generally unavailable without an analytic calculation or draws
from $P$, \fct{fssd\_null\_pvalue} instead estimates
$\Sigma_Q=\COV_{Q}\{\tau(X)\}$ from the observed feature
matrix.  Under $H_0$, $\Sigma_Q=\Sigma_P$, and this sample covariance
converges to their common value, giving an asymptotic level-$\alpha$ test.

\fct{fssd\_null\_pvalue} draws $B$ replicates
$S_{\tilde n}^{*(b)}=\sum_{q=1}^{dJ}\widehat\lambda_q
\{(Z_q^{(b)})^2-1\}$ from the null law in Equation~\ref{eq:fssd-null} and
returns a list whose \code{p_value} is
$\widehat p=B^{-1}\sum_{b=1}^B
\mathbf{1}\{S_{\tilde n}^{*(b)}\geq S_{\tilde n}\}$.
Here $B$ is set by the public \code{n\_simulations} argument.
\fct{fssd\_test} automatically constructs the feature matrix, evaluates the
statistic, and performs this null simulation.  It reports $S_{\tilde n}$ as
\code{fssd}.

\subsubsection{Random test locations} \label{sec:fssd-rand}

FSSD-rand is available as \fct{fssd\_rand\_test} or as
\code{fssd_test(..., variant = "rand")}.  It fits a Gaussian distribution to
\code{X}, draws $J$ locations from it, and uses all rows to compute the
statistic and its plug-in null distribution.  It requires neither a split nor
numerical optimization.  The same observations are therefore used to select
the test locations and, when needed, the kernel scale, and to compute the test
statistic and its null approximation.  These selections are not independent of
the tested observations, as required by the stated null calibration, so the
reported $p$~value is not guaranteed to have nominal level.  The method
nevertheless has a population justification: Theorem~1 of
\citet{Jitkrittum+Xu+Szabo+Fukumizu+Gretton:2017} shows that random locations
drawn from a distribution with a density yield a population FSSD that is zero
if and only if $Q=P$, almost surely over the locations.  Thus, FSSD-rand
retains population identification provided that the fitted Gaussian is
nondegenerate and the conditions below hold, but its plug-in calibration is
heuristic.

The population identification result requires the following conditions
\citep[Theorem~1]{Jitkrittum+Xu+Szabo+Fukumizu+Gretton:2017}: The
state space is connected and open; the kernel is real analytic and
$C_0$-universal; the moment and tail conditions of Theorem~1 hold; and the
locations have a Lebesgue density.  The fixed-location null calibration
additionally requires the second-moment condition of Proposition~2 and
independence of $L$ and the kernel scale from the i.i.d.\ test rows.  The
Gaussian RBF satisfies the kernel conditions, whereas IMQ and custom kernels
require separate verification.

The example supplies a fixed scale through \code{gof_kernel}:
\begin{CodeChunk}
\begin{CodeInput}
R> set.seed(1103)
R> fit_r <- fssd_test(X, score_p, variant = "rand", J = 2,
+    kernel = gof_kernel, n_simulations = 9999)
R> fit_r
\end{CodeInput}
\begin{CodeOutput}
Finite Set Stein Discrepancy (rand) - gaussian_rbf

data:  X
fssd = -0.10064, n_simulations = 9999, J = 2, scaling = 1, p-value = 0.4999
\end{CodeOutput}
\end{CodeChunk}
The plug-in test does not reject $P$ ($S_{\tilde n}=-0.10064$, $p=0.4999$).
Its off-diagonal estimator can be negative in finite samples.

\subsubsection{Optimized test locations} \label{sec:fssd-opt}

FSSD-opt is available through \fct{fssd\_opt\_test}.  It is also the default
\code{variant = "opt"} of \fct{fssd\_test}.  Unlike FSSD-rand, which tests on
all rows, it divides them into disjoint training and test sets of sizes
$n_{\mathrm{train}}$ and $n_{\mathrm{test}}$.  On the training rows it chooses
the locations $L$.  When the kernel has an adjustable scale, it chooses that
scale as well.  The test rows are reserved for the final statistic and null
calibration.  It selects these
parameters by maximizing the following approximate power criterion:
\begin{equation} \label{eq:fssd-power-criterion}
\frac{\widehat{\mathrm{FSSD}}^2}
     {\sqrt{\widehat{\VAR}_{H_1}}+\gamma},
\qquad
\widehat{\VAR}_{H_1}
=4\bar\tau^\top\widehat\Sigma_\tau\bar\tau,
\qquad \gamma>0.
\end{equation}
The training-sample plug-in quantities are
\[
\bar\tau=n_{\mathrm{train}}^{-1}\sum_{i=1}^{n_{\mathrm{train}}}\tau_i,
\qquad
\widehat\Sigma_\tau=n_{\mathrm{train}}^{-1}
\sum_{i=1}^{n_{\mathrm{train}}}(\tau_i-\bar\tau)(\tau_i-\bar\tau)^\top.
\]
Proposition~4 of
\citet{Jitkrittum+Xu+Szabo+Fukumizu+Gretton:2017} shows that, for a large test
sample, power is governed by $\mathrm{FSSD}^2/\sigma_{H_1}$.  The criterion
above is its plug-in version, so larger values indicate higher approximate
power.  The large-sample approximation assumes
$\sigma_{H_1}^2=4\mu_{L,h^2}^\top\Sigma_{Q,L,h^2}\mu_{L,h^2}>0$, with
$\mu_{L,h^2}$ and $\Sigma_{Q,L,h^2}$ the population counterparts of $\bar\tau$
and $\widehat\Sigma_\tau$ at the candidate locations and scale (the null
calibration above uses the same covariance at the values finally selected).
The fixed constant $\gamma>0$ keeps the empirical denominator away from zero.

The criterion in Equation~\ref{eq:fssd-power-criterion} is optimized by a
bounded quasi-Newton search, which needs
starting values and bounds.  A Gaussian fit to the training rows initializes the
locations, within the training range extended by \code{locs_bounds_frac}
standard deviations.  If the kernel has an adjustable scale, its squared scale
is optimized jointly with the locations, within \code{scale_lower} and
\code{scale_upper},
starting from a short grid around a median-based training scale when none is
supplied and from the supplied value otherwise.  The search has
$dJ+1$ parameters when the kernel scale is adjustable and $dJ$ otherwise.  The
package warns when this count exceeds $n_{\mathrm{train}}$.

The optimized locations and any optimized kernel scale are then fixed, and
Equations~\ref{eq:fssd-tau}--\ref{eq:fssd-null} are evaluated only on the
held-out rows.  Conditional on the training rows, the selected locations and
any selected kernel scale are fixed independently of the test rows, so
Equation~\ref{eq:fssd-null} applies as stated.
The split therefore restores the guarantee that FSSD-rand cannot provide:
\begin{CodeChunk}
\begin{CodeInput}
R> set.seed(1108)
R> fit_o <- fssd_opt_test(X, score_p, J = 2,
+    kernel = gof_kernel, train_ratio = 0.2, maxit = 50,
+    n_simulations = 9999)
R> fit_o
\end{CodeInput}
\begin{CodeOutput}
Finite Set Stein Discrepancy (opt) - gaussian_rbf

data:  X
fssd = -0.015262, n_simulations = 9999.00000, J = 2.00000, train_ratio =
0.20000, gamma = 0.00010, scaling = 0.73791, p-value = 0.7654
\end{CodeOutput}
\end{CodeChunk}
\code{train_ratio = 0.2} uses $80$ rows for optimization and $320$ for the
final test, so the reported \code{fssd} is
$S_{320}=320\widehat{\mathrm{FSSD}}^2$ on those held-out rows.  The test does not
reject $P$.  The example starts at the $h^2=1$ of \code{gof_kernel} and
returns $h^2=0.73791$.

The criterion is generally non-concave and may have several local maxima, so the
returned locations and any optimized scale may depend on the starting values and
optimization bounds.  The procedure is not guaranteed to find a global maximum,
and the optimizer may return a solution on the boundary.  This limitation does
not apply to FSSD-rand, which involves no numerical optimization.  The returned
locations and any optimized scale are reported, since they jointly define the
fitted test, and \code{fit_o\$info} records the optimization results and
diagnostics.  The package fixes no random number stream of its own, so
\fct{set.seed} before the call is what makes the split, the starting
locations, and the null simulation reproducible.

\section{Sampling and compression} \label{sec:sampling}

The task is to choose $m$ equally weighted points $x_1,\ldots,x_m$ whose
empirical distribution $Q_m$ approximates $P$ without requiring its
normalizing constant or exact draws.  All four methods use the score $s_p$,
while their additional inputs differ.  \fct{svgd} transports a fixed point set
along the direction of steepest local decrease in Kullback--Leibler divergence \citep{Liu+Wang:2016}.  The other
three share a greedy KSD objective and differ mainly in how candidates are
supplied \citep{Chen+Mackey+Gorham+Briol+Oates:2018,
Chen+Barp+Briol+Gorham+Girolami+Mackey+Oates:2019,
Riabiz+Chen+Cockayne+Swietach+Niederer+Mackey+Oates:2022}.
Table~\ref{tab:sampling} gives each method's candidate source and the inputs it
needs beyond the score.

\begin{table}[!ht]
\centering
\footnotesize
\setlength{\tabcolsep}{2pt}
\begin{tabular}{@{}llll@{}}
\hline
Function & Role & Update/candidates & Method-specific input \\
\hline
\fct{svgd} & transport &
all-point update &
initial points \\
\fct{stein\_points} & generation &
optimizer over $\mathcal X$ &
\code{log_p} or \code{x_init}; optimizer \\
\fct{sp\_mcmc} & generation &
short-chain states &
\code{log_p}, \code{x_init}; MCMC transition, \code{m_seq}, start rule \\
\fct{stein\_thinning} & compression &
rows of \code{X} &
\code{X}; \code{S} or \code{score_function} \\
\hline
\end{tabular}
\caption{\label{tab:sampling}The four sampling and compression methods of the
package.}
\end{table}

\FloatBarrier

\subsection{Common target distribution} \label{sec:sampling-interface}

As a running illustration, all four methods below approximate an equal mixture of two
bivariate normals with unit covariance and means $(\pm2,0)^\top$.  Two modes
expose a failure that a unimodal $P$ would hide.  A method that leaves all
its points in one mode still produces a tight point set, but it misses half of
$P$.  Each constructed output has $m=50$ points.  The code
also defines a fixed IMQ base kernel at $c=1$ and $\beta=-1/2$.  This exponent
is in the range where the associated Langevin KSD
controls weak convergence \citep[Theorem~8]{Gorham+Mackey:2017}.
\begin{CodeChunk}
\begin{CodeInput}
R> mu <- matrix(c(-2, 0, 2, 0), nrow = 2)
R> sigma <- array(diag(2), c(2, 2, 2))
R> target <- gmm(nComp = 2, mu = mu, sigma = sigma,
+    weights = c(0.5, 0.5), d = 2)
R> score_mix <- get_score_evaluator(target)
R> log_p_mix <- function(X) log(densitygmm(target, X))
R> kern <- stein_kernel("imq", c = 1, beta = -0.5)
\end{CodeInput}
\end{CodeChunk}
\code{score_mix} supplies the vectorized score $s_p$, and \code{log_p_mix}
the log density $\log p$, used later for the initial point of Stein Points and
for SP-MCMC.

\subsection{Transport by Stein variational gradient descent}
\label{sec:svgd}

SVGD, the transport method of Table~\ref{tab:sampling}, moves all $m$ points
at each iteration using a single map whose vector field is chosen as a local descent
direction toward $P$.  Writing $Q$ for the population distribution of the
current point set, SVGD chooses a vector field $\phi\in\mathcal H_k^d$ so that
the map $T_\epsilon(x)=x+\epsilon\phi(x)$ decreases
$\mathrm{KL}(Q_{[T_\epsilon]}\Vert P)$ as rapidly as possible, where
$Q_{[T_\epsilon]}$ is the distribution of $T_\epsilon(X)$ for $X\sim Q$.
For $Q$ with a density, that rate is
\begin{equation} \label{eq:svgd-kl}
\left.
\frac{d}{d\epsilon}
\mathrm{KL}(Q_{[T_\epsilon]}\Vert P)
\right|_{\epsilon=0}
=-\E_{X\sim Q}\{(\mathcal T_p\phi)(X)\}
\end{equation}
\citep[Theorem~3.1]{Liu+Wang:2016}.  Thus the steepest descent direction
maximizes this Stein expectation over the unit ball of $\mathcal H_k^d$.

Averaging Equation~\ref{eq:stein-feature} over $X\sim Q$ turns that expectation
into an inner product between $\phi$ and the Stein witness
\begin{equation} \label{eq:svgd-direction}
\phi^*_{Q,P}(\cdot)
=\E_{X\sim Q}\{\xi_p(X,\cdot)\}
=\E_{X\sim Q}\left\{
  k(X,\cdot)s_p(X)+\nabla k(X,\cdot)
 \right\},
\end{equation}
and that is the inner product maximized in Section~\ref{sec:kernelized} to
define $\mathrm{KSD}_{k,p}$.  Its maximum over the unit ball is
$\|\phi^*_{Q,P}\|_{\mathcal H_k^d}=\mathrm{KSD}_{k,p}(Q)$, attained in the
normalized direction of $\phi^*_{Q,P}$ when it is nonzero.  SVGD moves the
points along the unnormalized witness $\phi^*_{Q,P}$, and the step size $\epsilon$
controls the magnitude of the displacement.
Under the conditions of Section~\ref{sec:kernelized}, the witness is zero only
when $Q=P$.  This identifies $P$, but does not
guarantee convergence of the finite-point updates below.

Equation~\ref{eq:svgd-direction} is an expectation under $Q$, which the
algorithm evaluates at the points representing it, giving the update
\begin{equation} \label{eq:svgd-update}
x_i
\leftarrow
x_i+\frac{\epsilon}{m}\sum_{j=1}^{m}
 \left\{
  k(x_j,x_i)s_p(x_j)
  +\nabla_{x_j}k(x_j,x_i)
 \right\},
\qquad i=1,\ldots,m.
\end{equation}
The first term inside the braces moves points toward regions where $p$ is
larger, while the second separates nearby points to prevent them from
collapsing onto the same mode of $\log p$.

Applying Equation~\ref{eq:svgd-update} requires evaluating all point pairs, at
a cost of $O(m^2d)$ per update, and is implemented through
\code{svgd(x0, score_function, ...)}.  The $m\times d$ matrix \code{x0}
contains the initial points, and \code{score_function} evaluates
$s_p(x)$.  The arguments \code{n_iter} and \code{step_size}
set the number of updates and $\epsilon$.  \fct{svgd} returns an object whose
\code{X} field holds the final $m\times d$ point matrix.  With
\code{trace_iters}, \code{trace} additionally stores the post-update point
matrices at the requested iterations, named by iteration.

Write $g_t$ for the raw direction in Equation~\ref{eq:svgd-update} at
iteration $t$.  By default, the package rescales each entry by its running
root mean square,
\[
H_1=g_1^2,
\qquad
H_t=\kappa H_{t-1}+(1-\kappa)g_t^2,
\qquad
x
\leftarrow
x
+\epsilon\,\frac{g_t}{10^{-6}+\sqrt{H_t}},
\]
where operations are entrywise and $\kappa$, set by \code{alpha}, defaults to
$0.9$.  This
scaling is used in the original implementation of \citet{Liu+Wang:2016}.
If supplied, \code{adj_grad} is applied to the raw direction $g_t$ in place
of the default scaling $g_t/(10^{-6}+\sqrt{H_t})$.
Setting \code{adj_grad = function(grad, ...) grad} recovers
Equation~\ref{eq:svgd-update} as stated by \citet{Liu+Wang:2016}.

A call to \fct{svgd} uses one kernel family and its supplied parameters.
The default is the Gaussian RBF kernel of Equation~\ref{eq:rbf-kernel}, which
\citet{Liu+Wang:2016} use, rather than the IMQ kernel of
Section~\ref{sec:sampling-interface} used by the other three methods.  If the
Gaussian RBF bandwidth $h$ is not supplied, it
is recomputed from the current points at every iteration as
\[
h=\frac{\rho}{\sqrt{2\log m}},
\]
where $\rho$ is the current median pairwise distance.  The bandwidth therefore
tracks $\rho$ as the points move.\footnote{The package writes the RBF exponent with denominator $2h^2$.
This rule is therefore equivalent to $h_{\mathrm{LW}}=\rho^2/\log m$ in the
parameterization of \citet{Liu+Wang:2016}, whose bandwidth $h_{\mathrm{LW}}$
enters as $\exp(-\lVert x-y\rVert^2/h_{\mathrm{LW}})$.}  Only \fct{svgd}
recomputes a bandwidth this way.  The other three add points one at a time to
minimize a single KSD, so a
bandwidth that changed between steps would change what they minimize.

For the running mixture, the complete call is
\begin{CodeChunk}
\begin{CodeInput}
R> set.seed(2101)
R> x0 <- matrix(rnorm(100), ncol = 2)
R> particles <- svgd(x0, score_mix,
+    n_iter = 500, step_size = 0.1,
+    adj_grad = function(grad, ...) grad)$X
R> round(c(left_share = mean(particles[, 1] < 0),
+    left_mean = mean(particles[particles[, 1] < 0, 1]),
+    right_mean = mean(particles[particles[, 1] >= 0, 1])), 3)
\end{CodeInput}
\begin{CodeOutput}
left_share  left_mean right_mean
     0.520     -1.975      1.965
\end{CodeOutput}
\end{CodeChunk}
The call initializes $50$ points near the origin and returns their locations
after $500$ updates with the plain direction specified by \code{adj_grad}.
Of the final points, $52\%$ lie to the left of zero, with first-coordinate
means $-1.975$ on the left and $1.965$ on the right, so this run places them
in the two component regions.

\subsection{Greedy KSD construction} \label{sec:greedy-construction}

The remaining methods construct the empirical approximation one point at a
time.  After selecting $x_1,\ldots,x_{j-1}$, adding a candidate $x$ gives
\[
\tfrac12 j^2\mathrm{KSD}_j^2
=
\tfrac12\sum_{a=1}^{j-1}\sum_{b=1}^{j-1}k_{0,p}(x_a,x_b)
+\tfrac12 k_{0,p}(x,x)
+\sum_{i=1}^{j-1}k_{0,p}(x_i,x).
\]
The first term does not involve $x$, so minimizing the new discrepancy is
equivalent to
\begin{equation} \label{eq:sp-greedy}
x_j\in\arg\min_{x\in\mathcal A_j}
 \left\{
  k_{0,p}(x,x)+2\sum_{i=1}^{j-1}k_{0,p}(x_i,x)
 \right\}.
\end{equation}
Here $\mathcal A_j$ is the set of candidates available at step $j$.  For the
original greedy Stein Points method it is the whole domain of $p$,
$\mathcal A_j=\mathcal X=\mathbb R^d$, which is equivalent to Equation~(7) of
\citet{Chen+Mackey+Gorham+Briol+Oates:2018}.  SP-MCMC replaces
$\mathcal X$ by states visited by a short Markov chain
\citep[Section~3.1]{Chen+Barp+Briol+Gorham+Girolami+Mackey+Oates:2019},
whereas Stein Thinning replaces it by rows of an existing sample
\citep[Algorithm~1]{Riabiz+Chen+Cockayne+Swietach+Niederer+Mackey+Oates:2022}.
Among the greedy versions of the three methods, the common objective is fixed
and the candidate source is the central distinction.

\subsubsection{Stein points: Numerical global search}
\label{sec:stein-points}

Stein Points applies Equation~\ref{eq:sp-greedy} with $\mathcal A_j=\mathcal X$, so at
every step the objective is minimized over the whole domain.  At step $j=1$,
nothing has been selected yet, so the interaction sum
$\sum_{i<j}k_{0,p}(x_i,x)$ has no terms and the self term
$k_{0,p}(x,x)$ is the whole objective.  Two initialization rules
appear in existing work: \citet[Section~3.1]{Chen+Mackey+Gorham+Briol+Oates:2018}
suggest a global maximum of $p$, which needs no normalizing constant, while the
released \proglang{MATLAB} code minimizes the self term
\citep{Chen+EtAl:SteinPointsSoftware}.  \fct{stein\_points} uses
\code{x_init} directly when it is supplied.  Otherwise, it uses the same
\code{optimizer} used in later steps to find an approximate maximizer of
\code{log_p}.  The package does not compute the \proglang{MATLAB} start,
so users who want it must set \code{x_init} to an approximate minimizer of
$k_{0,p}(x,x)$.  The \code{x_init} argument also fixes the first point in
\fct{sp\_mcmc}.

Minimizing over $\mathcal X$ is usually nonconvex with no closed form, so each
step is solved approximately by scoring a finite set of candidate rows.  The
package separates these two roles.  \fct{stein\_points} holds the
selected points, their scores, and the Stein kernel,
and evaluates the objective $k_{0,p}(x,x)+2\sum_{i<j}k_{0,p}(x_i,x)$ for any matrix
of candidates.  The \code{optimizer} argument controls the candidate search.
It proposes rows at step $j$ and returns the best one, its score, its objective
value, and the number of rows evaluated, which accumulates in
\code{cum_n_eval}.  Because the optimizer does not access the kernel, any
function with the same arguments and return value can replace it, and the
helpers below return optimizer functions rather than points.

The three built-in optimizer constructors follow Algorithms~1--3 of
\citet{Chen+Mackey+Gorham+Briol+Oates:2018}.  They differ mainly in how they
propose candidates.  \fct{fmin\_grid} places
a regular grid over the search box and does not adapt to the selected points.
At step $j$, \fct{fmin\_grid} scores a Cartesian grid
with $n_0+\mathrm{round}(\sqrt{j})$ nodes along each dimension,
with $\{n_0+\mathrm{round}(\sqrt{j})\}^d$ candidates in total, so it is practical
mainly for small $d$.  \fct{fmin\_mc} instead scores \code{n_mc} random draws,
using a fixed Gaussian for the first \code{delay} steps and thereafter an
equally weighted Gaussian mixture centered at the selected points, with
component covariance \code{sigsq}$I$.  \fct{fmin\_nm} draws from that
same proposal, but treats its \code{n_res} draws as starting points rather than
candidates.  Nelder--Mead descends from each, and the best local minimum is
returned.  All searches are constrained to the box defined by \code{lb} and
\code{ub}, so that box, not $\mathcal X$, is the domain actually searched.

For the running mixture, multi-start Nelder--Mead within a box
containing both modes gives
\begin{CodeChunk}
\begin{CodeInput}
R> set.seed(2102)
R> opt <- fmin_nm(lb = c(-6, -4), ub = c(6, 4))
R> sp <- stein_points(score_mix, kern, n_points = 50, d = 2,
+    optimizer = opt, log_p = log_p_mix)
R> round(sp$ksd[c(1, 10, 50)], 3)
\end{CodeInput}
\begin{CodeOutput}
[1] 1.414 0.365 0.105
\end{CodeOutput}
\end{CodeChunk}
The output reports the KSD after selecting 1, 10, and 50 points.  It
decreases from 1.414 to 0.105 in this run.  The selected points are stored in
\code{sp\$X}.  For the first $j$ points, \code{sp\$ksd[j]} and
\code{sp\$cum_n_eval[j]} give the KSD and cumulative evaluation count,
respectively.

Under the kernel and per-step optimization conditions of
\citet[Theorem~2]{Chen+Mackey+Gorham+Briol+Oates:2018}, approximate greedy
selection has KSD $O\{\sqrt{\log(m)/m}\}$.  A uniform per-step
suboptimality bound $\delta$ enters the squared bound as $\delta/m$, so the
optimizer must control that error.

\paragraph{Coordinate refinement.} Once \fct{stein\_points} selects a point,
later steps leave it fixed.  \fct{stein\_codescent} instead refines a completed
point set without changing its size.  It takes each point in turn, holds all
other points fixed, and searches for a replacement.  It accepts the replacement
only if the KSD does not increase
\citep[Section~3.1]{Chen+Mackey+Gorham+Briol+Oates:2018}.

\paragraph{Herding and truncation.}
\code{method = "herding"} drops the self term from selection, while the
reported diagnostic remains the full KSD.  Without that term, the convergence
guarantee in \citet[Theorem~1]{Chen+Mackey+Gorham+Briol+Oates:2018} requires
selected candidates to lie in $\{x:k_{0,p}(x,x)\leq R_j^2\}$.  The arguments
\code{c2} and \code{truncation} set this bound or switch it off.  The theorem's
\code{c2}
depends on $P$ and $k_{0,p}$ and is generally unknown.  The package does not
infer it, so users must supply a value and assess sensitivity.  The filter
starts at the second point, so any supplied \code{x_init} must be checked
separately.

\paragraph{Alternative kernels.}
\begin{sloppypar}
Because the Stein Points selection objective depends on the base kernel,
\citet{Chen+Mackey+Gorham+Briol+Oates:2018} considered two alternatives to the
standard position-based IMQ kernel.  The package implements them through
\fct{stein\_kernel\_inverse\_log} and \fct{stein\_kernel\_imq\_score}:
\begin{align*}
k_{\mathrm{IL}}(x,y)
&=\{\alpha+\log(1+\lVert x-y\rVert^2)\}^{\beta},
&&\alpha>0,\quad\beta<0,\\
k_{\mathrm{score}}(x,y)
&=\{\alpha+\lVert s_p(x)-s_p(y)\rVert^2\}^{\beta},
&&\alpha>0,\quad\beta\in(-1,0).
\end{align*}
The inverse-log kernel measures Euclidean separation and retains slowly
decaying interactions between distant points, whereas the score-based kernel
compares the target scores $s_p(x)$ and $s_p(y)$ and therefore requires
\code{hess_log_p} to evaluate its derivatives.
\end{sloppypar}

\subsubsection{SP-MCMC: Markov-chain candidates} \label{sec:sp-mcmc}

SP-MCMC replaces the global search over $\mathcal X$ with a local one.  At step
$j$, it runs a short Markov chain and takes the states that chain visits as the
candidate set.  \fct{sp\_mcmc} requires \code{x_init}, which fixes $x_1$.
Each chain starts from an already selected point and its repeated states are
dropped.  The remaining states are scored by Equation~\ref{eq:sp-greedy}, and
the minimizer becomes $x_j$
\citep[Section~3.1]{Chen+Barp+Briol+Gorham+Girolami+Mackey+Oates:2019}.

\code{criterion} decides which selected point a chain starts from.
\code{"last"} uses $x_{j-1}$, \code{"rand"} draws uniformly from
$x_1,\ldots,x_{j-1}$, and \code{"infl"} uses the point whose removal maximizes
the remaining-set KSD.  Custom start rules can also be supplied through
\code{criterion}.

\code{mcmc} selects \fct{rwm} or \fct{mala}.  Random-walk Metropolis (RWM)
proposes $Y\mid x\sim\mathcal N(x,\eta\Sigma)$.  The Metropolis-adjusted
Langevin algorithm (MALA) adds the score drift and proposes
$Y\mid x\sim\mathcal N\{x+(\eta/2)\Sigma s_p(x),\eta\Sigma\}$ before applying
the Metropolis correction
\citep[Section~3.1]{Chen+Barp+Briol+Gorham+Girolami+Mackey+Oates:2019}.
\code{transition_fn} replaces the transition, and
\code{proposal_fn} sets the step size $\eta$, supplied as \code{h}, and
$\Sigma$ per step.

\code{m_seq} sets the chain length $\nu_j$, which counts every state including
the starting point
\citep[Section~3.1]{Chen+Barp+Briol+Gorham+Girolami+Mackey+Oates:2019}, so the
chain makes $\nu_j-1$ transitions and \code{m_seq = 1} makes none.  A scalar
\code{m_seq} uses the same chain length at every point-selection step, whereas a vector
assigns a separate length to each step.  Longer chains generate more
states at higher transition cost, but not necessarily more distinct candidates.

For the running mixture, the following call fixes $x_1$ at the left
component mean and constructs the other 49 points from MALA chains of length
ten:
\begin{CodeChunk}
\begin{CodeInput}
R> set.seed(2103)
R> spm <- sp_mcmc(score_mix, log_p_mix, kern, n_points = 50, d = 2,
+    mcmc = "mala", criterion = "last", m_seq = 10, h = 0.5,
+    x_init = c(-2, 0))
R> round(spm$ksd[c(1, 10, 50)], 3)
\end{CodeInput}
\begin{CodeOutput}
[1] 1.414 0.411 0.145
\end{CodeOutput}
\end{CodeChunk}
\begin{CodeChunk}
\begin{CodeInput}
R> c(sp_evals = sp$cum_n_eval[50],
+    spm_evals = spm$cum_n_eval[50],
+    accept = round(mean(spm$accept_rate, na.rm = TRUE), 3))
\end{CodeInput}
\begin{CodeOutput}
 sp_evals spm_evals    accept
 4821.000   981.000     0.964
\end{CodeOutput}
\end{CodeChunk}
Compared with the Stein Points run of Section~\ref{sec:stein-points}, the KSD
is higher ($0.145$ versus $0.105$) but is obtained from 981 rather than 4821
evaluations of $p$: one initial score evaluation plus 49 chains of ten
log-density and ten score evaluations.  MALA scores each chain state as it is generated, so candidate
selection reuses those scores.  The totals mix log-density and score
evaluations and are not runtimes.

On average, the MALA chains accepted $96.4\%$ of proposed moves (\code{accept}).
\code{counts} separates log-density, score, and
candidate-score evaluations that \code{cum_n_eval} aggregates.  If a chain
never moves, the step re-selects its starting point.  Repeated selections are
summarized by \code{n_repeated}.
The user can increase \code{m_seq} or retune the transition or proposal.

For convergence in expected squared KSD under Theorem~2 of
\citet{Chen+Barp+Briol+Gorham+Girolami+Mackey+Oates:2019},
use a kernel and a Markov transition satisfying its assumptions,
keep the transition fixed across point-selection steps, and choose
\code{m_seq} so that the bound tends to zero as more points are selected.

\subsubsection{Stein thinning: An existing candidate set}
\label{sec:stein-thinning}

Stein Thinning compresses a sample that already exists.  Given rows
$z_1,\ldots,z_n$ and their scores, it minimizes Equation~\ref{eq:sp-greedy}
over those $n$ rows at every step and returns $m$ indices
$\pi(1),\ldots,\pi(m)$
\citep[Algorithm~1]{Riabiz+Chen+Cockayne+Swietach+Niederer+Mackey+Oates:2022}.
The package maintains the objective in Equation~\ref{eq:sp-greedy} for each
input point.  At every step, it selects the point with the smallest objective
value among all $n$ input points, then updates the objectives by evaluating
the Stein kernel between the selected point and all $n$ input points.
Selecting $m$ points therefore requires $O(nm)$ Stein-kernel evaluations
\citep[cf.][Remark~2]{Riabiz+Chen+Cockayne+Swietach+Niederer+Mackey+Oates:2022}.
The same input index may be selected more than once.  If index $i$ appears $c$
times in \code{idx}, then the input state \code{X[i, ]} receives mass $c/m$ in
the empirical measure of the thinned output.  Consequently, \code{X[idx, ]} may
contain repeated rows.  Thus, Stein Thinning reweights states in the input
sample but does not generate new states.

\fct{stein\_thinning} needs the sample \code{X} and either a score matrix
\code{S} or a \code{score_function}.  It returns \code{idx}, so that
\code{X[idx, ]} is the compressed sample.  When \code{S} is supplied,
\mbox{\code{S[idx, ]}} contains the corresponding scores without rescoring.
For a fixed positive-definite
$\Gamma$, \citet{Chen+Barp+Briol+Gorham+Girolami+Mackey+Oates:2019} proposed
the preconditioned IMQ KSD below to improve performance:
\begin{equation} \label{eq:thin-kernel}
k(x,y)=\left\{1+\lVert\Gamma^{-1/2}(x-y)\rVert^2\right\}^{-1/2},
\qquad \Gamma_{\mathrm{med}}=\rho^2I,
\qquad \Gamma_{\mathrm{sclmed}}=\frac{\rho^2}{\log(m)}I.
\end{equation}
Here $\rho$ is the median pairwise distance, whereas \code{"smpcov"} uses the
sample covariance of all rows.  The two median rules, with \code{"sclmed"} the
default, use rows selected according to \code{pre_subsample}.  By default, these
are the first $\min(n,1000)$ rows, as in the paper
\citep[Section~2.3]{Riabiz+Chen+Cockayne+Swietach+Niederer+Mackey+Oates:2022}.
The default \code{"sclmed"} requires $m>1$.
Both median rules require at least two preconditioning rows, and
\code{"smpcov"} requires a nonsingular sample covariance.
$\Gamma^{-1}$ corresponds to the \code{precon} parameter in
Section~\ref{sec:gof-kernels},
and \fct{stein\_thinning} sets it from \code{pre} and warns if the supplied kernel
already carries one.

The following call thins a 2000-state random-walk chain to 50 points.  For
comparison, the full chain and thinned sample are evaluated by the root
V-statistic KSD of Section~\ref{sec:ksd-v}, using the fixed \code{kern} rather
than the preconditioned objective of \fct{stein\_thinning}:
\begin{CodeChunk}
\begin{CodeInput}
R> set.seed(2104)
R> chain <- rwm(log_p_mix, x0 = c(-8, -6), h = 0.3, Sigma = diag(2),
+    m_iter = 2000)
R> idx <- stein_thinning(chain$X, score_function = score_mix, m = 50)

R> root_ksd <- function(X, S, kernel) {
+    sqrt(ksd_v_statistic(stein_kernel_matrix(kernel, X, S)) / nrow(X))
+  }
R> round(c(ksd_chain = root_ksd(chain$X, score_mix(chain$X), kern),
+    ksd_thin = root_ksd(chain$X[idx, ], score_mix(chain$X[idx, ]), kern),
+    left_chain = mean(chain$X[, 1] < 0),
+    left_thin = mean(chain$X[idx, 1] < 0)), 3)
\end{CodeInput}
\begin{CodeOutput}
 ksd_chain   ksd_thin left_chain  left_thin
     0.157      0.098      0.324      0.520
\end{CodeOutput}
\end{CodeChunk}
The chain started at $(-8,-6)^\top$ and then oversampled the right mode,
leaving $32\%$ of its states in the left half-plane.  The 50 selected rows place
$52\%$ of their mass in the left half-plane, against a probability of $50\%$
under $P$, and lower the KSD from $0.157$ to $0.098$.

\section{Case studies} \label{sec:application}

We reproduce two real-data applications from
\citet{Jitkrittum+Xu+Szabo+Fukumizu+Gretton:2017} and
\citet{Chen+Barp+Briol+Gorham+Girolami+Mackey+Oates:2019} and extend each on the same data
to show how to apply the package and interpret its results.\footnote{Some
original data and implementation details were unavailable, so minor differences
may remain.}

\subsection{Model criticism for Chicago robbery locations}
\label{sec:app-chicago}

We revisit the Chicago robbery-location application of
\citet{Jitkrittum+Xu+Szabo+Fukumizu+Gretton:2017} using the 2016 robbery
records, each represented by a longitude--latitude pair.\footnote{Data:
Chicago Police Department's \emph{Crimes -- 2001 to Present} portal
\citep{ChicagoPD:2026}.}
We ask whether a Gaussian mixture fitted to part of the data represents the
distribution of robbery locations.  For each fitted model $P$, we test
$H_0:Q=P$, where $Q$ is the distribution generating the locations, and examine
the FSSD power criterion $C$ to identify regions of mismatch.

\begin{figure}[!ht]
\centering
\includegraphics[width=0.60\textwidth]{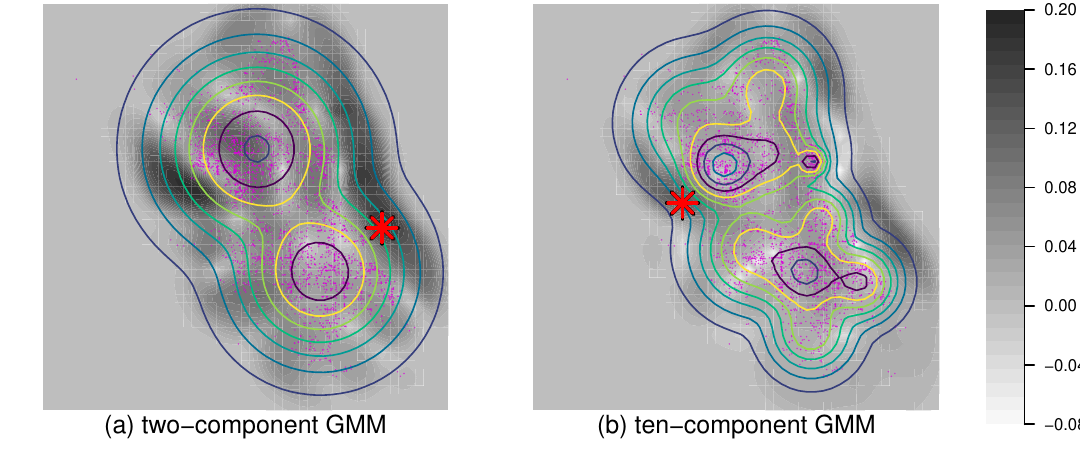}
\caption{\label{fig:app-gof}Reweighted mixtures with (a) $M=2$ and (b) $M=10$.}
\end{figure}

We fit $P$ on $5500$ of the selected $11{,}000$ records, independently of
the observations used for testing, as required in
Section~\ref{sec:gof-interface}.  $X_{\mathrm{tune}}$ contains $1100$
observations for tuning the FSSD test location and bandwidth, while
$X_{\mathrm{test}}$ contains $4400$ observations for evaluation.
FSSD calibration requires the selected location and
bandwidth to be independent of the test observations, as explained in
Section~\ref{sec:fssd-opt}.  This motivates separate tuning and test sets,
although the source-study bandwidth initialization uses both sets.  For
Figure~\ref{fig:app-gof}, $X_{\mathrm{plot}}$ contains $2750$ rows sampled
from their combined $5500$ rows.  It overlaps these sets and is used only for
visualization.\looseness=-1

We fit spherical Gaussian mixtures with $M=2$ and $M=10$ components to
check whether increasing the number of components yields an adequate fit.
We reproduce the source study's data splits and fits in \proglang{Python}
\citep{Python,Harris+EtAl:2020,Pedregosa+EtAl:2011,pandas} using its reported
seeds and fit the mixtures with
\code{GaussianMixture(covariance_type = "spherical", random_state = 3)}.
Each component has a fitted mean, covariance $s_j^2 I_2$, and weight $w_j$.\looseness=-1

The \proglang{Python} preparation script exports the data splits and fitted
parameters, which are read into \proglang{R} as \code{X\_tune},
\code{X\_test}, \code{X\_plot}, and \code{gmm\_parameters}.
We pass these fitted parameters to \fct{gmm} and obtain the score with
\fct{get\_score\_evaluator}.  The following code constructs the ten-component
model from \code{gmm\_parameters}.  The columns of \code{mu10} are the component
means, the slices of \code{sigma10} are their covariance matrices, and
\code{weights10} contains their mixture weights.
\begin{CodeChunk}
\begin{CodeInput}
R> parameters10 <- gmm_parameters[
+    gmm_parameters$n_components == 10, , drop = FALSE]
R> parameters10 <- parameters10[order(parameters10$component), ]
R> mu10 <- rbind(
+    longitude = parameters10$longitude_mean,
+    latitude = parameters10$latitude_mean)
R> sigma10 <- array(0, dim = c(2, 2, 10))
R> sigma10[1, 1, ] <- sigma10[2, 2, ] <- parameters10$variance
R> weights10 <- parameters10$fit_weight
R> model10 <- gmm(nComp = 10, mu = mu10, sigma = sigma10,
+    weights = weights10, d = 2)
R> score10 <- get_score_evaluator(model10)
\end{CodeInput}
\end{CodeChunk}
Here \fct{gmm} constructs \code{model10} from the fitted parameters,
and \fct{get\_score\_evaluator} returns its score function \code{score10}.
The tests access the fitted model only through this score function.
\code{score2} is obtained similarly for $M=2$.
Users can apply the same workflow to another model by supplying their
data and the corresponding score function.

Following the source study, we select $J=1$ location per model by maximizing
$C$ to identify local mismatch under the testing problem above.
We use the Gaussian RBF kernel supported by the FSSD theory in
Section~\ref{sec:fssd-rand}.\footnote{The source study uses
$C(L,h^2;\lambda)=\widehat{\mathrm{FSSD}}^2/
\{\widehat{\VAR}_{H_1}+\lambda\}^{1/2}$, where $\lambda>0$ keeps the denominator away from zero.
This differs from the criterion in \fct{fssd\_opt\_test}
(Equation~\ref{eq:fssd-power-criterion}), so we implement it using package functions.}

\begin{sloppypar}
\interlinepenalty=10000
\fct{source\_power\_objective} evaluates $C$ at given \code{L} and \code{h2}
by combining \fct{stein\_kernel},
\fct{compute\_tau}, and \fct{fssd\_statistic};
\code{sigma\_h1\_sq} estimates $\VAR_{H_1}$, and \code{reg} is $\lambda$.
\end{sloppypar}
\begin{CodeChunk}
\begin{CodeInput}
R> source_power_objective <- function(X, score_fn, L, h2, reg) {
+    kernel_obj <- stein_kernel("gaussian_rbf", h = sqrt(h2))
+    tau <- compute_tau(X, score_fn(X), L, kernel_obj)
+    mu <- colMeans(tau)
+    sigma_h1_sq <- 4 * mean(as.numeric(tau %*% mu)^2) - 4 * sum(mu^2)^2
+    (fssd_statistic(tau) / nrow(tau)) /
+      sqrt(max(sigma_h1_sq + reg, .Machine$double.eps))
+  }
\end{CodeInput}
\end{CodeChunk}
\begin{sloppypar}
Optimization requires starting values for
$L$ and $h^2$.  We reproduce the source study's initialization by choosing a
random row of $X_{\mathrm{tune}}$ for $L$ and using
\fct{median\_h2\_subsample} on the combined tuning and test sets for $h^2$.
This function draws $1000$ rows with a fixed seed and computes
their median scale with \fct{find\_median\_distance}.
We clip the initial $h^2$ to $[7\times10^{-4},10^{-3}]$.  For new data,
choose the interval based on squared distances
between observations and check whether the optimized $h^2$ reaches either
endpoint, as described in Section~\ref{sec:fssd-opt}.
Our helper \fct{fit\_source\_fssd}
then runs a bounded quasi-Newton search to maximize $C$ over $L$ and $h^2$
on $X_{\mathrm{tune}}$.
It calls \fct{source\_power\_objective} to evaluate $C$ with $\lambda=0.1$.
We fix the returned $L$ and $h^2$ and compute the statistic
and null $p$~value
on $X_{\mathrm{test}}$:\looseness=-1
\end{sloppypar}
\begin{CodeChunk}
\begin{CodeInput}
R> fssd10_fit <- fit_source_fssd(score10)
R> tau10 <- compute_tau(X_test, score10(X_test), fssd10_fit$L,
+    stein_kernel("gaussian_rbf", h = sqrt(fssd10_fit$h2)))
R> statistic10 <- fssd_statistic(tau10)
R> set.seed(10)
R> fssd10_null <- fssd_null_pvalue(
+    tau10, statistic10, n_simulations = 9999)
\end{CodeInput}
\end{CodeChunk}
For $M=2$, replace \code{score10} with \code{score2}.
To reproduce the source-study figure, we multiply each mixture weight $w_j$
by its component standard deviation $s_j$ and renormalize.
This upweights components with larger standard deviations
in the plotted density without changing their means or variances.
The tests use the original weights.
We then evaluate $C$ on $X_{\mathrm{plot}}$
over a location grid, using a bandwidth retuned as above for each
reweighted mixture and $\lambda=0.01$.\looseness=-1
\FloatBarrier

In Figure~\ref{fig:app-gof}, gray shading shows the FSSD power criterion $C$,
and darker shading indicates a larger value.
Purple points are $X_{\mathrm{plot}}$, and colored contours show the mixture
density.  The red star, at the maximum of $C$, lies in the right tail for
$M=2$ and in the left region for $M=10$.  These locations do not rank the
overall fit of the two mixtures.  Comparing the points and density contours
near each star helps identify local mismatch and possible model revisions.

\paragraph{Further analysis.}
Under the i.i.d.\ observation assumption, we further analyze the overall fit of
the original mixtures using KSD-U and KSD-V, asking whether they also reject
$Q=P$.  Both tests use the same $4400$ rows of $X_{\mathrm{test}}$ and reuse
\code{score2} and \code{score10} as the model inputs.  We use the Gaussian
RBF kernel, supported by the KSD testing theory in Section~\ref{sec:ksd-tests},
with a shared bandwidth
chosen by \fct{median\_h2\_subsample} on $X_{\mathrm{tune}}$, keeping
bandwidth selection separate from testing:
\begin{CodeChunk}
\begin{CodeInput}
R> ksd_h2 <- median_h2_subsample(X_tune, size = 1000,
+    seed = 9827)
R> ksd_kernel <- stein_kernel("gaussian_rbf", h = sqrt(ksd_h2))
R> set.seed(2201)
R> ksd_u <- ksd_u_test(X_test, score10, kernel = ksd_kernel,
+    nboot = 9999)
R> set.seed(2202)
R> ksd_v <- ksd_v_test(X_test, score10, kernel = ksd_kernel,
+    boot_method = "rademacher", nboot = 9999)
\end{CodeInput}
\end{CodeChunk}

The $M=2$ runs replace \code{score10} by \code{score2}.
Table~\ref{tab:app-gof} also includes the FSSD-opt $p$~values computed above.

\begin{table}[!ht]
\centering
\small
\begin{tabular}{lrr}
\hline
$p$~value & $M=2$ & $M=10$\\
\hline
FSSD-opt &${<}0.0001$ & ${<}0.0001$\\
KSD-U     & 0.0001   & 0.0022\\
KSD-V     & 0.0001   & 0.0025\\
\hline
\end{tabular}
\caption{\label{tab:app-gof}Goodness-of-fit $p$~values on $X_{\mathrm{test}}$.}
\end{table}

\FloatBarrier
Both KSD tests also reject the original fitted
mixtures.  For further diagnosis, map $C$ using the original weights,
and assess any revised model on fresh test data.

\subsection{Posterior approximation and compression for an IGARCH model}
\label{sec:igarch}

Daily stock returns alternate between calm and turbulent periods.  An integrated
generalized autoregressive conditional heteroskedasticity (IGARCH) model
represents this volatility persistence by letting today's conditional
variance depend on yesterday's squared return and conditional variance.  For
returns $(y_t)$, the model \citep{Taylor:2011} is
\begin{equation} \label{eq:igarch}
y_t=\sigma_t\varepsilon_t,
\qquad
\sigma_t^2=\theta_1+\theta_2y_{t-1}^2+(1-\theta_2)\sigma_{t-1}^2,
\qquad
\varepsilon_t\stackrel{\mathrm{iid}}{\sim}\mathcal N(0,1).
\end{equation}
Here $\sigma_t^2$ is the conditional variance, $\theta_1>0$ is a positive
constant, and $\theta_2\in(0,1)$ weights the latest squared return, with
$\theta_2+(1-\theta_2)=1$ giving IGARCH its unit persistence.
We use the $2000$ daily S\&P~500 percentage returns from 6 December
2005 to 14 November 2013 studied by
\citet[Section~4.4]{Chen+Mackey+Gorham+Briol+Oates:2018}.

Let $\theta=(\theta_1,\theta_2)^\top\in
\mathcal X=\mathbb R^+\times(0,1)$.  Up to an additive constant, the conditional
Gaussian log-likelihood is
\begin{equation} \label{eq:igarch-loglik}
\ell(\theta)
=-\frac12\sum_{t=2}^{T}
\left\{\log\sigma_t^2(\theta)
+\frac{y_t^2}{\sigma_t^2(\theta)}\right\},
\qquad T=2000,
\end{equation}
where $\sigma_t^2(\theta)$ is generated recursively using
Equation~\ref{eq:igarch}, with the sample variance as the initial value.  With
the improper uniform prior on
$\mathcal X$ used by \citet{Chen+Mackey+Gorham+Briol+Oates:2018},
$p(\theta\mid y)\propto\exp\{\ell(\theta)\}\mathbf 1_{\mathcal X}(\theta)$.
The posterior places negligible mass near $\partial\mathcal X$, so, following
the original studies, we neglect the boundary term in this experiment
\citep{Chen+Mackey+Gorham+Briol+Oates:2018,
Chen+Barp+Briol+Gorham+Girolami+Mackey+Oates:2019}.

As in Section~\ref{sec:sampling}, the methods need only the score of this
posterior and, for the MCMC steps, its log density; users can apply the same
workflow to another posterior by supplying these two functions.  Since the prior is
constant on $\mathcal X$, the score is the gradient of the log-likelihood in
Equation~\ref{eq:igarch-loglik}:
\begin{equation*}
\nabla_\theta\log p(\theta\mid y)
=\sum_{t=2}^{T}\frac{y_t^2-\sigma_t^2}{2\sigma_t^4}\,\nabla_\theta\sigma_t^2,
\qquad
\nabla_\theta\sigma_t^2=(1,\,y_{t-1}^2-\sigma_{t-1}^2)^\top
+(1-\theta_2)\,\nabla_\theta\sigma_{t-1}^2,
\end{equation*}
where the recursion, obtained by differentiating Equation~\ref{eq:igarch},
starts from $\nabla_\theta\sigma_1^2=0$ because the initial variance is
fixed at the sample variance.

The following code binds \fct{igarch\_logp} and \fct{igarch\_score} to the
observed S\&P~500 returns $y_t$ in Equations~\ref{eq:igarch}--\ref{eq:igarch-loglik},
using their sample variance as the initial conditional variance $\sigma_1^2$.
The resulting functions \code{log_p(X)} and \code{score_fn(X)} evaluate the
log posterior and its score at the parameter pairs
$\theta=(\theta_1,\theta_2)^\top$ supplied as rows of \code{X}:
\begin{samepage}
\begin{CodeChunk}
\begin{CodeInput}
R> spx <- read.csv(file.path(replication_dir, "spx_returns.csv"))
R> rows <- spx$date >= "2005-12-06" & spx$date <= "2013-11-14"
R> returns <- spx$return_percent[rows]
R> initial_variance <- stats::var(returns)
R> log_p <- function(X) igarch_logp(X, returns, initial_variance)
R> score_fn <- function(X) igarch_score(X, returns, initial_variance)
\end{CodeInput}
\end{CodeChunk}
\end{samepage}
The same score function is used for the KSD evaluation in
Table~\ref{tab:igarch-thinning}.  We use this posterior to reproduce the
$1000$-point comparison of
\citet{Chen+Barp+Briol+Gorham+Girolami+Mackey+Oates:2019} and, as a further
analysis, to compress a $20{,}000$-state simulation to $100$ points.

\paragraph{Point construction.}
We compare nine methods.  Each produces $1000$ points that approximate the
posterior.
The MALA and RWM baselines each simulate $5000$ states and retain every fifth
state.  SVGD starts from $1000$ parameter values sampled from the ranges used
in the original study.  The remaining methods are minimum energy design (MED), Stein
Points, and four SP-MCMC configurations that combine a MALA or RWM transition
with the \code{"last"} or \code{"infl"} criterion.  MED is a comparator from
the original experiment rather than a package method \citep{Roshan+EtAl:2015}.

\begin{figure}[!ht]
\centering
\includegraphics[width=0.75\textwidth]{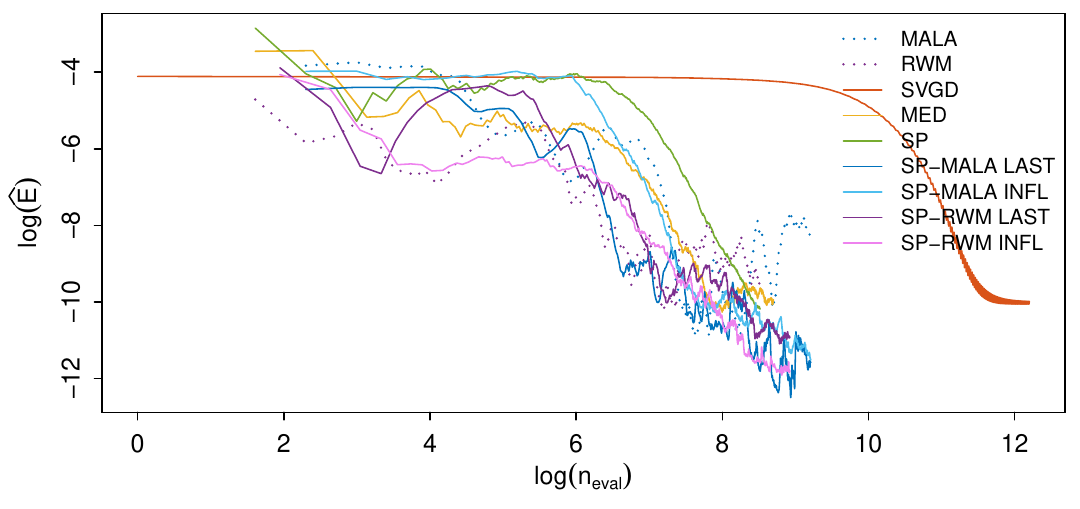}
\caption{\label{fig:igarch-energy}Energy distance to the
reference chain versus cumulative
evaluation count.}
\end{figure}

Because exact posterior draws are unavailable, a long MALA chain provides
an empirical reference, a covariance estimate for kernel scaling, and the
sample for compression.
We generate it with \fct{mala\_adapt}.  Shorter preliminary runs adjust the
proposal covariance to the posterior's spread and correlation and tune the
step size using the proportion of accepted moves.  We then hold these
settings fixed for a final $100{,}000$-state run and retain $20{,}000$ evenly
spaced states as \code{reference\_chain}.  The reported evaluation counts
exclude this preparation.  Figure~\ref{fig:igarch-energy} and
Table~\ref{tab:igarch-thinning} measure agreement with the
reference chain.

For approximation points $x_1,\ldots,x_m$ and reference states
$y_1,\ldots,y_N$, the empirical energy distance is
\[
\widehat{\mathcal E}
=\frac{2}{mN}\sum_{i=1}^{m}\sum_{j=1}^{N}\lVert x_i-y_j\rVert
-\frac{1}{m^2}\sum_{i=1}^{m}\sum_{i'=1}^{m}\lVert x_i-x_{i'}\rVert
-\frac{1}{N^2}\sum_{j=1}^{N}\sum_{j'=1}^{N}\lVert y_j-y_{j'}\rVert,
\]
including the diagonal pairs $i=i'$ and $j=j'$
\citep{Szekely+Rizzo:2004,Baringhaus+Franz:2004}.  Smaller values indicate a
closer approximation to the reference distribution.

To compare computational effort, Figure~\ref{fig:igarch-energy} plots energy
distance against the cumulative number of log-density and score evaluations,
on natural-log axes.  We count evaluations at individual
parameter values, so a call on $b$ points counts as $b$ evaluations.  MALA uses
one log-density and one score evaluation per simulated state, whereas RWM uses
one log-density evaluation.  Each SVGD update scores all $1000$ points and
therefore contributes $1000$ evaluations
\citep[Section~4.3]{Chen+Barp+Briol+Gorham+Girolami+Mackey+Oates:2019}.
For SP-MCMC, the count also includes candidate scores not already computed
by the transition.  These counts measure function evaluations rather than
elapsed time.  At the same evaluation count, the methods may represent the
posterior with different numbers of points.  SVGD evaluates energy distance
using $1000$ points throughout, whereas the other methods use the points accumulated so far, up to $1000$.

\raggedbottom
\begin{samepage}
We use the source-study settings with \code{V\_MCMC} $=5\times10^{-4}I_2$.
The MALA and RWM
baselines start at \code{X0} $=(0.021,0.125)^\top$ and retain every fifth
state (\code{N\_THIN} $=5$).
The calls are:\footnote{\code{thin\_idx} selects every fifth row
of each chain's \code{X} matrix.  The function \fct{nearest\_spd} ensures
that the proposal covariance is numerically positive definite.}
\begin{CodeChunk}
\begin{CodeInput}
R> set.seed(2101)
R> n_states <- 1000 * N_THIN
R> mala_chain <- mala(log_p, score_fn, X0, h = 0.13^2,
+    Sigma = V_MCMC, m_iter = n_states)
R> thin_idx <- seq.int(N_THIN, n_states, by = N_THIN)
R> set.seed(2102)
R> rwm_chain <- rwm(log_p, X0, h = 1, Sigma = nearest_spd(0.06 * V_MCMC),
+    m_iter = n_states)
\end{CodeInput}
\end{CodeChunk}
\end{samepage}
\begin{sloppypar}
Following the IMQ preconditioning in Section~\ref{sec:stein-thinning}, we
scale the kernel shared by the Stein methods using the inverse of
\code{reference\_covariance}, the sample covariance of \code{reference\_chain}.
SVGD starts from $1000$ points drawn uniformly on
$[0.002,0.04]\times[0.05,0.2]$, with bounds \code{LB} and \code{UB}.
We run $199$ updates with step size $\epsilon=0.001$, applying the default
AdaGrad-style rescaling to the direction in Equation~\ref{eq:svgd-update}
(Section~\ref{sec:svgd}), and store the points after each update in
\code{trace}:
\end{sloppypar}
\begin{CodeChunk}
\begin{CodeInput}[samepage=true]
R> set.seed(2103)
R> particles0 <- cbind(
+    stats::runif(1000, LB[1], UB[1]),
+    stats::runif(1000, LB[2], UB[2])
+  )
\end{CodeInput}
\end{CodeChunk}
\begin{CodeChunk}
\begin{CodeInput}[samepage=true]
R> kernel <- stein_kernel("imq", c = 1, beta = -0.5,
+    precon = solve(reference_covariance))
\end{CodeInput}
\end{CodeChunk}
\begin{CodeChunk}
\begin{CodeInput}[samepage=true]
R> svgd_fit <- svgd(particles0, score_fn, kernel = kernel,
+    n_iter = 199, step_size = 0.001, trace_iters = seq_len(199))
\end{CodeInput}
\end{CodeChunk}
Following Section~\ref{sec:stein-points}, we define \fct{source\_optimizer}
to select the valid candidate with the smallest objective value among five
candidate points.  At each step, we either draw all five from
$\mathcal N(X_0,0.5V_{\mathrm{MCMC}})$, or sample five selected points with
replacement and add independent noise from
$\mathcal N(0,5\times10^{-6}I_2)$ to each.  The probability of choosing the
second rule increases with the number of selected points.  The same search
selects \code{sp\_x1} as the initial point by minimizing $k_{0,p}(x,x)$.
The argument \code{d = 2} specifies the dimension of $\theta$:
\begin{CodeChunk}
\begin{CodeInput}
R> sp <- stein_points(score_fn, kernel, n_points = 1000,
+    d = 2, optimizer = source_optimizer, x_init = sp_x1)
\end{CodeInput}
\end{CodeChunk}
We compute the MED comparator with \fct{med\_greedy\_path}, using \code{log_p}
and the same five-candidate \fct{source\_optimizer}.
For SP-MCMC, we define \fct{run\_source\_sp\_mcmc} to prepare the initial
point and adaptive proposal, call \fct{sp\_mcmc} to construct the point set,
and add the initial-search cost to the evaluation counts.
It chooses \code{x\_init} by minimizing $k_{0,p}(x,x)$ over the distinct states
of a five-state MALA or RWM chain, and defines \code{proposal\_fn} to mix
the initial covariance with the covariance of the current points.
\par
\begin{samepage}
In its internal call below, \code{mcmc}, \code{criterion}, \code{h}, and
\code{initial\_covariance} come from the four configurations that follow;
\code{x\_init} and \code{proposal\_fn} are constructed within the function,
and \code{M\_SPMCMC} $=5$ is the chain length:
\par
\begin{CodeChunk}
\begin{CodeInput}
R> fit <- sp_mcmc(
+    score_fn, log_p, kernel, n_points = 1000, d = 2,
+    mcmc = mcmc, criterion = criterion, m_seq = M_SPMCMC,
+    h = h, Sigma = initial_covariance, x_init = x_init,
+    proposal_fn = proposal_fn
+  )
\end{CodeInput}
\end{CodeChunk}
\end{samepage}
\pagebreak[0]
\begin{samepage}
We combine two transitions, MALA and RWM, with the \code{"last"} and
\code{"infl"} start rules from Section~\ref{sec:sp-mcmc}.  The two MALA
settings keep the product of $h$ and the initial covariance fixed, while the
RWM settings scale the empirical covariance by $2.38^2/d$ with $d=2$:
\begin{CodeChunk}
\begin{CodeInput}
R> sp_mala_last <- run_source_sp_mcmc(
+    "mala", "last", h = 0.5^2,
+    initial_covariance = 0.02 * V_MCMC / 0.5^2, seed = 2106)
R> sp_mala_infl <- run_source_sp_mcmc(
+    "mala", "infl", h = 0.8^2,
+    initial_covariance = 0.02 * V_MCMC / 0.8^2, seed = 2107)
\end{CodeInput}
\end{CodeChunk}
\end{samepage}
\pagebreak[0]
\begin{samepage}
\begin{CodeChunk}
\begin{CodeInput}
R> sp_rwm_last <- run_source_sp_mcmc(
+    "rwm", "last", h = 1,
+    initial_covariance = 0.2 * V_MCMC, seed = 2108,
+    covariance_scale = 2.38^2 / 2)
\end{CodeInput}
\end{CodeChunk}
\end{samepage}
\pagebreak[0]
\begin{samepage}
\begin{CodeChunk}
\begin{CodeInput}
R> sp_rwm_infl <- run_source_sp_mcmc(
+    "rwm", "infl", h = 1,
+    initial_covariance = 0.2 * V_MCMC, seed = 2109,
+    covariance_scale = 2.38^2 / 2)
\end{CodeInput}
\end{CodeChunk}
\end{samepage}

Figure~\ref{fig:igarch-energy} reproduces the source study's qualitative
result, but variability across seeds and reference chains is not assessed.  At roughly
$5000$--$10{,}000$ evaluations, the four SP-MCMC endpoints have lower energy
distances than Stein Points, MED, RWM, and MALA.  SVGD uses $199{,}000$ score
evaluations, so it is worthwhile mainly when \code{score\_fn} is cheap for
many points at once.  To choose a method, compare the curves at the evaluation
budget the application can afford.

\FloatBarrier
\paragraph{Further analysis.}
If evaluating quantities such as conditional volatility at all $20{,}000$
states of \code{reference\_chain} is too costly, we ask whether $100$ points
can represent the chain accurately enough for these calculations.

We compare Stein Thinning with regular thinning, random subsampling, and
Support Points.  Stein Thinning selects states from the reference chain by
minimizing KSD.  Support Points starts from sampled chain states and adjusts
their locations to reduce energy distance; we implement it in \fct{sp\_ccp}
\citep{Mak+Joseph:2018}.

We measure how well each $100$-point set represents the chain in two ways,
matching the two column groups of Table~\ref{tab:igarch-thinning}; for all
six measures, smaller is better.  First, three discrepancies measure how far
the points are from the target distribution.  The energy distance (ED)
compares them with the full chain.  The two KSDs compare them with the
posterior; both are root V-statistics computed by \fct{root\_ksd} with the
IMQ kernel of Section~\ref{sec:stein-thinning}, and they differ only in
preconditioning.  KSD$_{\mathrm{med}}$ uses the \code{"med"} preconditioner
estimated from the first $1000$ reference states, and KSD$_{\mathrm{smpcov}}$
uses the inverse sample covariance of all $20{,}000$ states, the same
\code{precon} as Stein Thinning.

Second, three relative errors measure how well the points reproduce
posterior summaries of the full chain: the mean $\mu$, the covariance
$\Sigma$, and the $0.9$ quantile $q_{0.9}$ of the day-$2000$ conditional
volatility $\sigma_{2000}(\theta)$.  With $\widehat\mu$, $\widehat\Sigma$, and
$\widehat q_{0.9}$ denoting the same quantities computed from the $100$
points, the errors are
\predisplaypenalty=0%
\[
e_\mu=\max_r\frac{|\widehat\mu_r-\mu_r|}{|\mu_r|},\qquad
e_\Sigma=\frac{\|\widehat\Sigma-\Sigma\|_F}{\|\Sigma\|_F},\qquad
e_q=\frac{|\widehat q_{0.9}-q_{0.9}|}{q_{0.9}}.
\]
\predisplaypenalty=10000%
We construct the four point sets as follows:
\begin{CodeChunk}
\begin{CodeInput}
R> reference_scores <- score_fn(reference_chain)
R> idx_stein <- stein_thinning(reference_chain, S = reference_scores,
+    m = 100, pre = "smpcov")
R> idx_regular <- round(seq(1, nrow(reference_chain), length.out = 100))
R> set.seed(4104)
R> idx_random <- sort(sample.int(nrow(reference_chain), 100))
R> set.seed(4105)
R> support_points <- sp_ccp(reference_chain, n = 100)
\end{CodeInput}
\end{CodeChunk}
\begin{table}[!htbp]
\centering
\scriptsize
\setlength{\tabcolsep}{3pt}
\begin{tabular}{@{}lrrrrrr@{}}
\hline
& \multicolumn{3}{c}{Discrepancy} &
  \multicolumn{3}{c}{Relative error} \\
\cline{2-4}\cline{5-7}
Method & KSD$_{\mathrm{med}}$ & KSD$_{\mathrm{smpcov}}$ & ED &
  Mean & Covariance & Volatility\\
\hline
Stein Thinning    & 9.961  & 14.97 & $2.994\times10^{-5}$ & 0.0055 & 0.0213 & 0.0008\\
Regular thinning  & 30.390 & 45.50 & $2.224\times10^{-4}$ & 0.0048 & 0.3467 & 0.0057\\
Random subsample  & 30.121 & 58.88 & $3.613\times10^{-4}$ & 0.0270 & 0.0582 & 0.0026\\
Support Points    & 5.379  & 17.95 & $7.185\times10^{-6}$ & 0.0002 & 0.0072 & 0.0004\\
\hline
\end{tabular}
\caption{\label{tab:igarch-thinning}$100$-point approximations of the
$20{,}000$-state reference chain.}
\end{table}

\FloatBarrier

In Table~\ref{tab:igarch-thinning}, Support Points has the lowest ED,
KSD$_{\mathrm{med}}$, and summary errors, and Stein Thinning has the lowest
KSD$_{\mathrm{smpcov}}$, the discrepancy it minimizes.  ED and the summary
errors need the full chain, available here only because it is the reference,
whereas the KSDs need only \code{score\_fn} and the points once their kernels
are fixed.  Support Points moves its points off the chain, so it suits analyses
that need only representative parameter values.  Stein Thinning keeps chain
states, so quantities computed along the chain can be reused.


\flushbottom
\section{Summary and discussion} \label{sec:summary}

The \pkg{steinsampling} package provides a unified \proglang{R} workflow for
Stein goodness-of-fit testing, point construction, and sample compression
through a shared score and kernel setup.  The current implementation
prioritizes flexibility and transparency over speed.  Multithreaded CPU or
GPU backends such as \pkg{RcppParallel} and \pkg{torch}
\citep{RcppParallel,torch} could accelerate it.
Random feature Stein discrepancies (R$\Phi$SDs) use importance-sampled Stein
features to give near-linear-time approximations without forming the full
Stein-kernel matrix \citep{Huggins+Mackey:2018}, and incomplete U-statistics
evaluate KSD-U on only a subset of observation pairs
\citep{Schrab+Kim+Guedj+Gretton:2022}.  Other extensions include additional
Stein operators and methods for constrained or discrete target distributions.


\bibliography{refs}



\end{document}